\documentclass[aps,prb,preprint,superscriptaddress,reprint,preprintnumbers,amssymb]{revtex4-1}

\usepackage{graphicx}
\usepackage[latin1]{inputenc}
\usepackage{braket}
\usepackage{pifont}
\usepackage{units}
\usepackage{wrapfig}
\usepackage{graphicx}
\usepackage{amsmath}
\usepackage{rotating}
\usepackage{color}
\usepackage{amsfonts}
\usepackage{amssymb}
\usepackage{moreverb}
\usepackage{url}
\usepackage{epsf}
\usepackage{multirow}
\usepackage{subfigure}
\usepackage{morefloats}
\usepackage{threeparttable}
\usepackage{cancel}
\usepackage{bm}
\usepackage{tensor}

\begin{document}

\title{Daubechies Wavelets for Linear Scaling Density Functional Theory}

\author{Stephan Mohr}
%\email{}
\affiliation{Institut f\"ur Physik, Universit\"at Basel, Klingelbergstr.\ 82, 4056 Basel, Switzerland}
\affiliation{Univ. Grenoble Alpes, CEA, INAC-SP2M, F-38000 Grenoble, France } 

\author{Laura E.\ Ratcliff}
%\email{}
\affiliation{Univ. Grenoble Alpes, CEA, INAC-SP2M, F-38000 Grenoble, France } 

\author{Paul Boulanger} 
%\email{paul.boulanger@cea.fr}
%\homepage{http://inac.cea.fr/L_Sim}
\affiliation{Univ. Grenoble Alpes, CEA, INAC-SP2M, F-38000 Grenoble, France } 
\affiliation{Institut N\'{e}el, CNRS and Universit\'{e} Joseph Fourier, B.P.\ 166, 38042 Grenoble Cedex 09, France} 

\author{Luigi Genovese}
%\email{}
\affiliation{Univ. Grenoble Alpes, CEA, INAC-SP2M, F-38000 Grenoble, France }

\author{Damien Caliste}
%\email{}
\affiliation{Univ. Grenoble Alpes, CEA, INAC-SP2M, F-38000 Grenoble, France }

\author{Thierry Deutsch}
%\email{}
\affiliation{Univ. Grenoble Alpes, CEA, INAC-SP2M, F-38000 Grenoble, France }

\author{Stefan Goedecker}
%\email{}
\affiliation{Institut f\"ur Physik, Universit\"at Basel, Klingelbergstr.\ 82, 4056 Basel, Switzerland}

\date{\today}

\begin{abstract}

	We demonstrate that Daubechies wavelets can be used to construct a minimal set of optimized localized adaptively-contracted basis functions in which the Kohn-Sham orbitals can be represented with an arbitrarily  high, controllable precision. Ground state energies and 
the forces acting on the ions can be calculated in this basis with the same accuracy 
as if they were calculated directly in a Daubechies wavelets basis, provided that the amplitude of these adaptively-contracted basis functions 
is sufficiently small on the surface of the localization region, which is guaranteed by the optimization procedure described in this work.
This approach reduces the computational costs of DFT calculations, and can be 
combined with sparse matrix algebra to obtain linear scaling with respect to the number of electrons in the system. 
Calculations on systems of 10,000 atoms or more thus become feasible in a systematic basis set with moderate computational resources. Further computational savings can be achieved by exploiting the similarity of the adaptively-contracted basis functions for closely related environments, e.g.\ in geometry optimizations or combined calculations of neutral and charged systems. 
\end{abstract}

%\pacs
%{
%}

\maketitle

\section{Introduction}

 The Kohn-Sham (KS) formalism of density functional theory (DFT)~\cite{hohenberg42,kohn43} is one of the most popular 
electronic structure methods  due to its good balance between accuracy and speed. 
Thanks to the development of new approximations to the exchange
correlation functional, this approach now allows many quantities (bond lengths, vibration frequencies, elastic
constants, etc.) to be calculated with errors of less than a few percent, which is sufficient for many
applications in solid state physics, chemistry, materials science, biology, geology and many other fields.
Although the KS approach has some shortcomings -- e.g.\ its inability to accurately describe the HOMO-LUMO separation or 
many-body (e.g.\ excitonic) effects, thus reducing its predictive power in the field of optics -- it has become 
the standard for the quantum simulation
of matter and also provides a well defined starting point for more accurate methods,
 such as the \textit{GW} approximation~\cite{Hedin19701}. 

Despite the efforts put forth to increase the efficiency of DFT calculations and the increasing computing
power of modern supercomputers, the applicability for standard calculations is limited to systems containing about a
thousand atoms, which is small compared to the size of systems of
interest in nanoscience. 
The reason for this is that standard electronic structure programs using systematic basis sets such as 
plane waves~\cite{abinit, 0953-8984-21-39-395502, 0953-8984-14-11-301} , finite elements~\cite{pask} or 
wavelets~\cite{genovese:014109} need a number of operations that scales as
the number of orbitals, $N_{orb}$, squared times the number of basis functions, $N_{basis}$, used to represent them.
Since both the
number of orbitals and the number of basis functions scale as the number of atoms, the overall cost scales as
$\mathcal{O}(N_{orb}^2N_{basis})=\mathcal O(N^3_{atom})$. 
Electronic structure programs that use Gaussians~\cite{NWchem} or atomic orbitals~\cite{aims} require in 
a standard implementation a matrix diagonalization which scales as $\mathcal O(N_{basis}^3)$.

To circumvent this problem, one can exploit Kohn's
nearsightedness principle~\cite{PhysRev.133.A171,PhysRevLett.76.3168,goedecker1998decay}, which states
that, for systems with a finite gap or for metals at finite temperature, all physical quantities 
are determined by the local environment. This is a consequence of the exponentially fast decay of
 the density matrix~\cite{cloizeaux1964energy,cloizeaux1964analytical,kohn1959analytic,baer1997sparsity,ismail-beigi1999locality,goedecker1998decay,he2001exponential}.
Therefore, it is theoretically possible to express the KS wavefunctions of a given system in terms of 
a minimal, localized basis set. In order to get highly accurate results while still keeping the size 
of the basis relatively small, such a basis has to 
depend on the local chemical environment. If this basis set were known or could be
approximated beforehand, it would lead to a computationally cheap tight-binding 
like approach~\cite{PhysRevB.58.7260,doi:10.1021/jp070186p}.
Of course, in practice it is not possible to determine this optimal localized basis set beforehand; 
instead it has to be built up iteratively during the calculation. 
This would result in $\mathcal{O}(N_{orb}^3)$ scaling, which is still equivalent to $\mathcal{O}(N^3_{atom})$, 
but with a much smaller prefactor than systematic approaches (e.g.\ plane waves) where the number of basis functions 
is far greater than the number of orbitals ($ N_{basis} \gg N_{orb}$).

However, the use of a strictly localized basis offers yet another possibility.
As has been demonstrated during the past twenty years~\cite{0034-4885-75-3-036503,RevModPhys.71.1085}, 
it is possible to truncate the density matrix and thus transform it
 into a sparse form 
by neglecting elements either when they are below a certain threshold, or
 when they correspond to localized orbitals which are too distant from each other.
This reduces the complexity of the algorithm to
$\mathcal{O}(N_{orb})$=$\mathcal{O}(N_{atom})$ 
and leads to so-called linear scaling (LS) DFT methods. 
Even though the exponential decay of the density matrix is as well present for metals at finite temperature, 
we will -- as most $\mathcal{O}(N_{atom})$ approaches -- focus on the simpler case of insulators.
Methods of this type have been implemented in numerous
codes such as  \textsc{onetep}~\cite{skylaris:084119}, Conquest~\cite{0953-8984-22-7-074207}, CP2K~\cite{cp2k} and
\textsc{siesta}~\cite{0953-8984-14-11-302}. Note, however, that the extent of the truncation 
impacts the accuracy due to the imposition of an additional constraint on the system, 
and is therefore left as a freely selectable parameter for the user.
This additional constraint also comes at the cost of extra computational steps, so that the
prefactor is greater than for standard DFT codes, even for a single iteration in the self-consistency cycle.
Furthermore, there can be problems with ill-conditioning when using strictly localized basis sets, which further increases the prefactor.
The combination of these two problems means that
for small systems the total calculation time is actually greater
when one imposes locality, but thanks to
the better scaling, there is a crossover point where the new algorithms become more efficient.

%\emph{in situ}
Our minimal set of localized adaptively-contracted basis functions, called support functions in the following, is obtained by an environment dependent optimization 
where the support functions are represented in terms of a fixed underlying wavelet basis set.
The term \textit{adaptively-contracted} should not be confused with the terminology of contracted basis functions often used in quantum chemistry, it is simply used to emphasize that there are two levels of basis functions, namely the underlying wavelets basis and the support functions which are built out of them.
%{In the language of quantum chemistry, these support functions could be denoted as environment dependent contracted wavelets.}
Because of the environment dependency, the size of this basis set is however for a given accuracy much smaller than the size of typical 
contracted Gaussian basis sets and we refer to this basis set therefore also as a minimal basis set. 

The choice of the underlying basis set is one of the most important aspects impacting the accuracy and efficiency of a linear scaling DFT code.
Ideally, it should feature compact support while still being orthogonal, thus allowing for a systematic convergence --
properties which are all offered by Daubechies wavelets basis sets~\cite{Debauchies}.  Furthermore, wavelets have built in multiresolution properties,
enabling an adaptive mesh with finer sampling close to the atoms where the most significant part of the orbitals is located; 
this can be particularly beneficial for inhomogeneous systems. Wavelets also have the distinct advantage 
that calculations can
be performed with all the standard boundary conditions -- free, wire, surface or periodic. 
This also means we can perform calculations on charged and polarized systems 
using free boundary conditions without the need for a compensating background charge.
It is therefore evident that the combination of the above features makes wavelets ideal for a LSDFT code.

% To exploit the above properties we have implemented a minimal basis method within
% the existing BigDFT code~\cite{genovese:014109}, which uses a 
% Daubechies wavelet basis.
% As will be shown, we use a set of quasi-orthogonal minimal basis functions which are generated under the 
% application of a confining potential which, on the one hand, ensures that they remain sufficiently localized,
%  and on the other hand removes the ill-conditioning.

This paper is organized as follows.  We first give an overview of the method, focussing in particular on 
the imposition of the localization constraint in Daubechies wavelets.
We then discuss the details, highlighting the novel features, following which we consider the calculation of atomic forces.  For this latter point, we demonstrate the remarkable 
result that, thanks to the compact support of Daubechies wavelets, the contribution of the Pulay-like forces, arising from the introduction of 
the localization regions, can be safely neglected in a typical calculation.
We then present results for a number of systems, illustrating the accuracy of the method for ground state energies and atomic forces.  We also demonstrate the improved scaling compared with standard BigDFT, showing that we are able to achieve linear scaling.  Finally, we highlight two cases where the minimal basis functions can be reused, resulting in further significant computational savings.

\section{Minimal adaptively-contracted basis}
\subsection{Kohn-Sham formalism in a minimal basis set}
The standard approach for performing Kohn-Sham DFT calculations is to calculate the Kohn-Sham orbitals  $\Ket{\varPsi_i}$
which satisfy the equation
\begin{eqnarray} \label{eq:ks}
\mathcal{H}_{KS}\Ket{\varPsi_i}=\varepsilon_i\Ket{\varPsi_i},
\end{eqnarray}
with
\begin{eqnarray}
\mathcal{H}_{KS}=-\frac{1}{2}\nabla^2 + \mathcal{V}_{KS}[\rho] + \mathcal{V}_{PSP}, %+V_{ext} %+V_H[\rho]+V_{XC}[\rho]+V_{PSP}+V_{ext},
\end{eqnarray}
where $\mathcal{V}_{KS}[\rho]$ contains the Hartree potential -- solution to the Poisson equation -- 
and the exchange-correlation
potential, while $\mathcal{V}_{PSP}$ contains the potential arising from the pseudopotential 
and the external potential created by
the ions. In the case of BigDFT, these are norm-conserving GTH-HGH~\cite{PhysRevB.58.3641} pseudopotentials and their
Krack variants~\cite{springerlink:10.1007/s00214-005-0655-y}, possibly with a nonlinear core correction~\cite{NLCCpaper}.
It is worth noting that the use of pseudopotentials does not only decrease the complexity of the
calculation by reducing the number of orbitals and avoiding the need of very high resolution around the nuclei,
but also offers the possibility of easily including relativistic effects.
Furthermore the calculation of the Hartree and exchange-correlation potentials are done in the same way as
in the original version of BigDFT~\cite{genovese:014109} and are thus not subject to any approximations.
For the exchange-correlation part we restrict ourselves to local functionals.

In our approach the KS orbitals are in turn expressed as a linear combination of support functions  $\ket{\phi_\alpha}$:
\begin{eqnarray}
 \Ket{\varPsi_i(\mathbf{r})}=\sum_\alpha c_i^\alpha\Ket{\phi_\alpha(\mathbf{r})}.
 \label{eq:expansion}
\end{eqnarray}
The density -- which can be obtained from the one-electron orbitals via $\rho(\mathbf{r})=\sum_i f_i|\varPsi_i(\mathbf{r})|^2 $, 
where $f_i$ is the occupation number of orbital $i$ -- is given by
\begin{eqnarray} 
 \rho(\mathbf{r}) = \sum_{\alpha,\beta}\phi_\alpha^*(\mathbf{r})K^{\alpha\beta}\phi_\beta(\mathbf{r}),
 \label{eq:density}
\end{eqnarray}
where $K^{\alpha\beta}=\sum_i f_i c_i^{*\alpha} c_i^\beta$ is the density
kernel. The latter is related
to the density matrix formulation of Hern\'andez and Gillan~\cite{hernandez_kernel},
since -- as follows from Eq.~\eqref{eq:expansion} --
\begin{equation}
 \begin{aligned}
 F(\mathbf{r},\mathbf{r}') &= \sum_i f_i \Ket{\varPsi_i(\mathbf{r})}\Bra{\varPsi_i(\mathbf{r}')} \\
   &= \sum_{\alpha,\beta} \Ket{\phi_\alpha(\mathbf{r})}K^{\alpha\beta}\Bra{\phi_\beta(\mathbf{r'})}\;.
 \end{aligned}
 \label{eq:density_matrix}
\end{equation}
Thus the density kernel is the representation of the density matrix in the support function basis.
We choose to have real support functions and thus from now on we will neglect the complex notation for this quantity.

The density matrix decays exponentially with respect to the distance $|\mathbf{r}-\mathbf{r}'|$ for systems with a 
finite gap or for metals at finite temperature~\cite{cloizeaux1964energy,cloizeaux1964analytical,kohn1959analytic,baer1997sparsity,ismail-beigi1999locality,goedecker1998decay,he2001exponential}.
In these cases it can therefore be represented by
strictly localized basis functions. A natural and exact choice for these 
would be the maximally localized Wannier functions 
which have the same exponential decay~\cite{PhysRevB.56.12847}.
Of course, these Wannier functions are not known beforehand. Therefore, in our case, the adaptively-contracted basis functions are
constructed \textit{in situ} during the self-consistency cycle 
and are expected to reach a quality similar to that of the exact Wannier functions.

In the formalism we have presented so far, the KS orbitals have to be optimized by minimizing the total energy with respect to the support functions and density kernel.
For a self-consistent calculation this is equivalent to minimizing the band structure energy, i.e.
\begin{eqnarray}
E_{BS}=\sum_{\alpha,\beta}K^{\alpha\beta}H_{\alpha\beta}\;,
\label{eq:band_structure_energy}
\end{eqnarray}
subject to the orthonormality condition of the KS orbitals,
\begin{eqnarray}\label{eq:ks_orthog}
\Braket{\varPsi_i|\varPsi_j}= \sum_{\alpha,\beta}c_i^{\alpha *}S_{\alpha\beta}c_j^\beta=\delta_{ij},
\label{eq:orth}
\end{eqnarray}
where $H_{\alpha\beta}=\Braket{\phi_\alpha|\mathcal{H}_{KS}|\phi_\beta}$ and $S_{\alpha\beta}=\Braket{\phi_\alpha|\phi_\beta}$ are the Hamiltonian and overlap matrices of the support functions, respectively.
For systems with all occupation numbers being either zero or one, this is equivalent to imposing the idempotency condition on the density kernel
$K^{\alpha \beta}$,
\begin{equation}\label{eq:idem}
\sum_{\gamma,\delta}K^{\alpha \gamma}S_{\gamma\delta}K^{\delta\beta} = K^{\alpha \beta},
\end{equation}
which can be achieved using
the McWeeny purification scheme~\cite{mcweeny41} or directly imposing the orthogonality constraint on the coefficients $c_i^\alpha$

\begin{figure}
 \centering
 \includegraphics[width=0.25\textwidth]{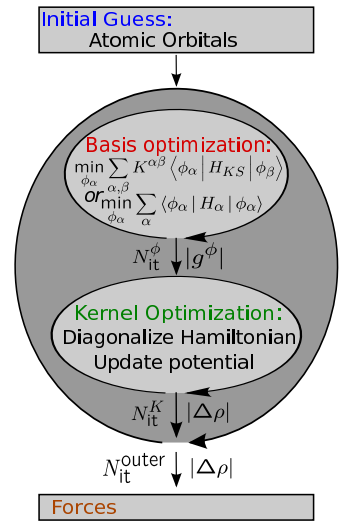}
 \caption{Structure of the minimal basis approach: for the basis optimization 
loop the hybrid scheme can be used instead of trace or energy minimization, and for the kernel optimization 
loop either direct minimization or the Fermi operator expansion method can be used in place of diagonalization; see Sec~\ref{sec:scc}.}
 \label{fig:overview}
\end{figure}

The algorithm therefore consists of two key components: support function and density kernel optimization. 
The workflow is illustrated in Fig.~\ref{fig:overview}; it consists of a flexible double loop 
structure, with the outer loop controlling the overall convergence, and two inner 
loops which optimize the support functions and density kernel, respectively. 
The first of these inner loops is done 
non-self-consistently (i.e.\ with a fixed potential), whereas the second one is done self-consistently.

\subsection{Daubechies wavelets in BigDFT}
\begin{figure}
\includegraphics[scale=0.35]{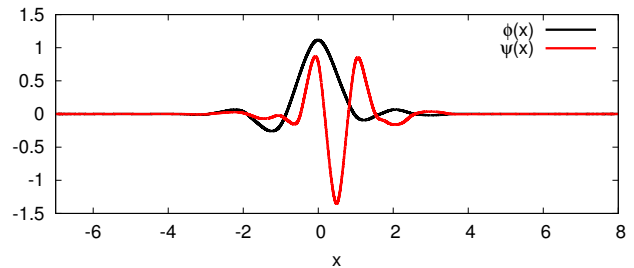}
\caption{Least asymmetric Daubechies wavelet family of order $2m=16$; both the scaling function $\phi(x)$ and
wavelet $\psi(x)$ differ from zero only within the interval $[1-m,m]$.\label{fig:Daub_16}}
\end{figure}

BigDFT~\cite{genovese:014109} uses the orthogonal least asymmetric Daubechies~\cite{Debauchies} family of order $2m=16$,
illustrated in Fig.~\ref{fig:Daub_16}. 
These functions have a compact support and are smooth, which means that they are also localized in
Fourier space. This wavelet family is able to exactly represent polynomials up to
$8^{\text{th}}$ order. Such a basis is therefore an optimal choice given that we desire at the same time locality and
interpolating power.
An exhaustive presentation of the use of wavelets in numerical simulations can be found in
Ref.~\onlinecite{goedecker1998wavelets}. 

A wavelet basis set is generated by the integer translates of the scaling functions and wavelets, with arguments
measured in units of the grid spacing $h$. In three dimensions, a wavelet basis set can easily be obtained as the
tensor product
of one-dimensional basis functions, combining wavelets and scaling functions along each coordinate of the Cartesian grid
(see \textit{e.g.}\ Ref.~\onlinecite{genovese:014109}).

In a simulation domain, we have three categories of grid points:
those which are closest to the atoms (``fine region'') carry one (three-dimensional) 
scaling function and seven (three-dimensional) wavelets; those which are further away 
from the atoms (``coarse region'') carry only one scaling function, corresponding to a 
resolution which is half that of the fine region; and those which are even 
further away (``empty region'') carry neither scaling functions nor wavelets.
The fine region is typically the region where chemical bonding takes place, whereas the 
coarse region covers the region where the tails of the wavefunctions decay smoothly to zero. 
We therefore have two resolution levels whilst maintaining a regularly spaced grid in the entire simulation box.

A support function $\phi_\alpha(\mathbf{r})$ can be expanded in this wavelet basis as follows:
\begin{eqnarray} \label{Eq:phi}
 \phi(\mathbf{r})&=&\sum_{i_1,i_2,i_3}s_{i_1,i_2,i_3}\varphi_{i_1,i_2,i_3}(\mathbf{r})\nonumber\\
&+&\sum_{j_1,j_2,j_3}\sum_{l=1}^{7}d_{j_1,j_2,j_3}^{(\ell)}\psi_{j_1,j_2,j_3}^{(\ell)}(\mathbf{r}),
\end{eqnarray}
where $\varphi_{i_1,i_2,i_3}(\mathbf{r})=\varphi(x-i_1) \varphi(y-i_2)
\varphi(z-i_3)$ is the tensor product of three one-dimensional scaling functions centered at the grid point $(i_1,i_2,i_3)$,
and $\psi_{j_1,j_2,j_3}^{(\ell)}(\mathbf{r})$ are the seven tensor products 
containing at least one one-dimensional wavelet centered on the grid point $(j_1,j_2,j_3)$.
The sums over $i_1$, $i_2$, $i_3$ ($j_1$, $j_2$, $j_3$) run over all grid points where scaling functions
(wavelets) are centered, i.e.\ all the points of the coarse (fine) grid.
The overall simulation box is chosen to be rectangular and is identical to the one in the standard version of BigDFT;
for simplicity the origin is chosen such that there are only positive grid coordinates, i.e.\ in a corner of the simulation domain.

To determine these regions of different resolution, we construct two spheres around each atom $a$; a small one
with radius $R_a^f=\lambda^f\cdot r_a^f$ and a large one with radius $R_a^c=\lambda^c\cdot r_a^c$ ($R_a^c>R_a^f$). The
values of $r_a^f$ and $r_a^c$ are characteristic for each atom type and are related to the covalent and van der Waals radii,
whereas $\lambda^f$ and $\lambda^c$ 
can be specified by the user in order to control the accuracy of the calculation. The fine (coarse) region is then
given by the union of all the small (large) spheres, as
shown in Fig.~\ref{fig: Locregs}. 
% Hence in BigDFT the basis set is controlled by three user specified parameters: systematic
% convergence of the total energy to the exact value for a fixed computational volume is achieved by decreasing the value of $h$,  
% the computational volume is determined by $\lambda^c$ and the convergence rate 
% is influenced by the choice of $\lambda^f$.

Hence in BigDFT the basis set is controlled by these three user specified parameters.  By reducing $h$ and/or increasing $\lambda^c$ and $\lambda^f$ the computational degrees of freedom are incremented, leading to a systematic convergence of the energy.

\section{Localization regions}\label{sec:locregs}

\begin{figure}
 \centering
 \includegraphics[width=0.4\textwidth]{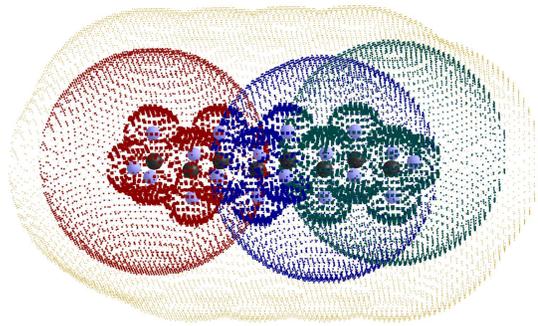}
 \caption{A two level adaptive grid for an alkane: the high resolution grid points are shown with bold points
while the low resolution grid points are shown with smaller points. Also visible are three localization regions (red, blue and
green) with radii of 3.7~\AA\ centered on different atoms in which certain support functions will reside.
The coarse grid points shown in yellow do not belong to any of the three localization regions.}
 \label{fig: Locregs}
\end{figure}

Thanks to the nearsightedness principle it is possible to define a  basis of strictly localized support functions such that the KS
orbitals given in terms of this adaptively-contracted basis are \textit{exactly} equivalent to the expression based solely
on the underlying Daubechies basis.
However, as presented so far, the support functions $\phi_\alpha(\mathbf{r})$ of Eq.~\eqref{Eq:phi} are expanded over the entire
simulation domain.
We want them to be strictly localized while still containing various resolution levels, as illustrated by Fig.~\ref{fig: Locregs}, and so we set to zero all scaling function and wavelet coefficients which
lie outside a sphere of radius $R_{cut}$ around the point $\mathbf{R}_\alpha$ on which the support function is centered.
In general, these centers $\mathbf{R}_\alpha$ could be anywhere, but we choose them to be centered on an atom $a$ 
and we thus assume from now on that $\mathbf{R}_\alpha = \mathbf{R}_a$.
Consequently we define a localization projector $\mathcal L^{(\alpha)}$, which is written in
the Daubechies basis space as
\begin{equation}
\mathcal L^{(\alpha)}_{i_1,i_2,i_3;j_1,j_2,j_3} = \delta_{i_1j_1} \delta_{i_2 j_2} \delta_{i_3j_3}
\theta(R_{cut} - |\mathbf{R}_{(i_1,i_2,i_3)}-\mathbf{R}_\alpha|)\;,
\end{equation}
where $\theta$ is the Heaviside function.  We use this projector to constrain the function $|\phi_\alpha \rangle$ to be localized throughout the calculation, i.e.
\begin{equation}\label{eq:loccon}
 |\phi_\alpha \rangle = \mathcal L^{(\alpha)} | \phi_\alpha \rangle\;.
\end{equation}
Clearly, if $| \phi_\alpha \rangle$ is localized around $\mathbf{R}_\alpha$ and $R_{cut}$ is large enough, $\mathcal
L^{(\alpha)}$ leaves  $| \phi_\alpha \rangle$  unchanged and no approximation is introduced to the KS equation.

% The support function can be represented in a compressed form where only the
% nonzero scaling function and wavelet coefficients are stored. Several operations such as scalar products among different
% orbitals and between orbitals and the nonlocal pseudopotential projectors can be done directly in this
% compressed form.

It is important to note that the localization constraint of Eq.~\eqref{eq:loccon} determines the expression of $\frac{\mathrm{d} \phi_\alpha}{\mathrm{d} \mathbf{R}_\beta}$ \textit{outside} the localization region of $ \phi_\alpha$.
Indeed, as the Daubechies basis set is independent of $\mathbf{R}_\alpha$, differentiating Eq.~\eqref{eq:loccon} with respect to $\mathbf{R}_\beta$ leads to
\begin{equation}\label{externalderivative}
 \left(1 - \mathcal L^{(\alpha)} \right) | \frac{\mathrm{d} \phi_\alpha}{\mathrm{d} \mathbf{R}_\beta} \rangle =
\delta_{\alpha\beta}
\frac{\partial \mathcal L^{(\alpha)}}{\partial \mathbf{R}_\alpha} | \phi_\alpha \rangle \;.
\end{equation}
This result will be used in Appendix~\ref{app:The case of the minimal basis setup} to demonstrate that the Pulay-like forces are negligible 
for a typical calculation with our approach.

\subsection{Imposing the localization constraint}\label{sec:imposingloc}

In what follows, we demonstrate that choosing the support functions to be orthogonal allows for a more straighforward 
application of the localization constraint.
Due to the orthonormality of the KS orbitals we cannot directly minimize the band structure energy (Eq.~\eqref{eq:band_structure_energy}) with respect to the support functions, rather we have to minimize the following functional:
\begin{equation}\label{eq:bstarget}
 \Omega = \sum_{\alpha,\beta}K^{\alpha\beta}H_{\alpha\beta} - 
\sum_{i,j}\sum_{\alpha,\beta} \Lambda_{ij} \left(c_i^{\alpha *} c_j^\beta  S_{\alpha\beta} - \delta_{ij}\right)\;,
\end{equation}
with the Lagrange multiplier coefficients $\Lambda_{ij}$ determined by the relation
\begin{equation}
 \sum_{i,j} c_i^{\alpha *}  c_j^\beta \Lambda_{ij} =
\sum_{\rho, \sigma} K^{\alpha \rho} H_{\rho\sigma} (S^{-1})^{\sigma \beta} \;,
\end{equation}
where $(S^{-1})^{\alpha \beta}$ is the inverse overlap matrix.
The gradient $ \Ket{g^\alpha}=\Ket{\frac{\delta \Omega}{\delta \Bra{\phi_\alpha}}}$ is therefore
\begin{eqnarray}
 \Ket{g^\alpha} &=& \sum_{\beta} K^{\alpha \beta} \mathcal{H}_{KS} \Ket{\phi_\beta} -
\sum_{\beta, \rho, \sigma} K^{\alpha \rho} H_{\sigma\rho}
(S^{-1})^{\sigma \beta}\Ket{\phi_\beta}. \label{unlocalizedgradient}
\end{eqnarray}
However, we wish to impose the localization condition  $|\phi_\alpha \rangle = \mathcal L^{(\alpha)} |
\phi_\alpha \rangle$ on the support functions and therefore 
the functional to be minimized becomes
\begin{equation}
\Omega'= \Omega - \sum_\alpha \Braket{\phi_\alpha| 1 - \mathcal L^{(\alpha)} | \ell^\alpha }\;,
\end{equation}
where the components of the vector $\Ket{ \ell^\alpha }$ are the Lagrange multipliers of this locality constraint.
The gradient for $\Omega'$, $\Ket{\frac{\delta \Omega'}{\delta \Bra{\phi_\alpha}}}$, can therefore be written as
\begin{equation}\label{eq:gradprime}
\Ket{g'^\alpha}=\Ket{g^\alpha}-(1 - \mathcal L^{(\alpha)})\Ket{ \ell^\alpha} \;.
\end{equation}
Using the stationarity condition $0=\Ket{g'^\alpha}$ and combining with the fact that $ 1 - \mathcal L^{(\alpha)}$ is a projection operator,
i.e.\ $(1 - \mathcal L^{(\alpha)})^2=1 - \mathcal L^{(\alpha)}$, we have
\begin{eqnarray}
(1 - \mathcal L^{(\alpha)})\Ket{ \ell^\alpha}=(1 - \mathcal L^{(\alpha)})\Ket{g^\alpha}\;.
\end{eqnarray}
Therefore, using Eq.~\eqref{eq:gradprime},
\begin{eqnarray}
\Ket{g'^\alpha}=\mathcal L^{(\alpha)}\Ket{ g^\alpha}\;.
\end{eqnarray}
i.e.\ the gradient is explicitly localized.
This yields the following result for the gradient:
\begin{eqnarray}
% \Ket{g'^\alpha} &=& \sum_{\beta} K^{\alpha \beta} \mathcal L^{(\alpha)} \mathcal{H}_{KS} \Ket{\phi_\beta}\nonumber\\
%&&-\sum_{\beta, \rho, \sigma} K^{\alpha \rho} H_{\sigma\rho}
%(S^{-1})^{\sigma \beta} \mathcal L^{(\alpha)} \Ket{\phi_\beta} \\
\Ket{g'^\alpha} &=& \sum_{\beta, \rho} K^{\alpha \rho} \tensor{(S^{1/2})}{_{\rho}^{\beta}}\bigg[ \mathcal L^{(\alpha)} \mathcal{H}_{KS}
\Ket{\widetilde \phi_\beta} \nonumber\\
&&- \sum_{\sigma}\Braket{\widetilde \phi_\sigma | \mathcal{H}_{KS} |\widetilde
\phi_\rho} \mathcal L^{(\alpha)} \Ket{\widetilde\phi_\sigma} \bigg]
\;.
\end{eqnarray}
Here the localized gradient is expressed in terms of the orthogonalized support functions 
$\Ket{\widetilde \phi_\alpha} = \sum_\beta \tensor{(S^{-1/2})}{_{\alpha}^{\beta}} \Ket{\phi_\beta}$.
Requiring the support functions to be orthogonal i.e.\ $S_{\alpha\beta}=\delta_{\alpha\beta}$ therefore further simplifies the evaluation of the gradient
as it no longer becomes necessary to calculate $\mathbf{S}^{-1}$ or $\mathbf{S}^{1/2}$.
Moreover, it avoids the need for distinguishing between covariant and contravariant indices~\cite{alvaro_chris}.

\subsection{Localization of the Hamiltonian application}
As shown in Ref.~\onlinecite{genovese:014109}, the Hamiltonian operator in a Daubechies wavelets basis set is
defined by a set of convolution operations, combined with the application of nonlocal pseudopotential projectors.
The nature of these operations is such that $\mathcal H_{KS}  | \phi_\beta \rangle$ will have a greater extent than $| \phi_\beta \rangle$.
We therefore define a second localization operator, $\mathcal L'^{(\beta)}$, with a corresponding cutoff radius $R_{cut}'$, such
that $R_{cut}'$ is equal to $R_{cut}$ plus half of the convolution filter length times the grid spacing, which in our case corresponds to an additional eight
grid points.  When applying the Hamiltonian, we impose
\begin{equation}
\mathcal H_{KS}| \phi_\beta \rangle = \mathcal L'^{(\beta)} \mathcal H_{KS} \mathcal L^{(\beta)} | \phi_\beta \rangle \;.
\end{equation}
For both the convolution operations and the nonlocal pseudopotential applications, this procedure guarantees that the Hamiltonian application is exact within the localization region of $\mathcal L^{(\beta)}$.
However, the values of $\mathcal H_{KS} \mathcal L^{(\beta)} | \phi_\beta \rangle$ are approximated outside this region due to the semilocal nature of the convolutions and the pseudopotential projectors. This impacts the evaluation of the Hamiltonian matrix $H_{\alpha\beta}$ for all elements whose localization regions do not coincide, and thus also affects the gradient $\Ket{g^\alpha}$.
Nonetheless, we have verified that further enlargement of $R'_{cut}$ is not needed as it has negligible impact on the accuracy, while adding additional overheads, see Sec.\ref{sec:results}.
%This is exact for the convolutions, but represents a small approximation for the nonlocal pseudopotential term.
%In practice we have seen that this is a reasonable approximation.

Apart from these technical details, most of the basic operations are identical to their
implementation in the standard BigDFT code~\cite{genovese:014109} and are therefore not repeated here. The only
difference is that these operations are now done only in the localization regions (corresponding to either $\mathcal L^{(\beta)}$ or $\mathcal L'^{(\beta)}$) and not 
in the entire computational volume.

%It important to emphasize that $\mathcal H_{KS}$ does not commute with the $\mathcal L^{(\alpha)}$ operators.
%In other terms the hamiltonian application will provide a results which is defined in 
%\begin{equation}
%\mathcal H_{KS}  | \phi_\beta \rangle  = \mathcal H_{KS} \mathcal L^{(\beta)} | \phi_\beta \rangle\neq 
%\mathcal L^{(\beta)} \mathcal H_{KS} | \phi_\beta \rangle.
%\end{equation}
%Therefore in order to calculate $ \mathcal H_{KS}  | \phi_\beta \rangle$ the
%localization constraint on the support functions has to be relaxed when applying the Hamiltonian
%operator. 
%These buffers are initialized to zero, but the convolution will result in non-zero values in
%those regions, which will affect scalar products with other support functions, i.e.\ the elements of the Hamiltonian matrix.
%The inclusion of the buffer regions is therefore crucial for the correct evaluation of the Hamiltonian matrix
%and thus for preserving the variationality of the procedure.
%When the scalar product with another basis
%function is evaluated, it is therefore important to keep this buffer zone.
%Thus the KS Hamiltonian can be explicitly evaluated within
%the truncation scheme applied, 
%preserving variationality of the result.

\section{Self-Consistent cycle}\label{sec:scc}
\subsection{Support function optimization}\label{sec:supportfunc}

As an initial guess for the support functions we use atomic orbitals, which are generated by solving the
atomic Schr\"odinger equation and therefore possess long tails which need to be truncated at the borders of the
localization regions. If the values at the borders are not negligible, the resulting kink will cause the kinetic energy to become very large
due to the definition of the Laplacian operator in a wavelet basis set, and so to assure stability during the optimization procedure,
the localization regions would need to be further enlarged.
To overcome this problem, even for small localization regions, it is advantageous to decrease the extent of the
atomic orbitals before the initial truncation by adding a confining quartic potential centered on each atom,
$a_{\alpha}(\mathbf{r}-\mathbf{R}_{\alpha})^4$, to the atomic Schr\"odinger equation. 

For the first few iterations of the outer loop (Fig.~\ref{fig:overview}) we maintain the quartic confining potential of the atomic input
guess. This implies that the total
Hamiltonian becomes dependent on the support function, 
$\mathcal{H}_\alpha = \mathcal{H}_{KS}+a_\alpha(\mathbf{r}-\mathbf{R}_\alpha)^4$, and we can no longer
minimize the band structure energy~(Eq.~\eqref{eq:band_structure_energy}) to obtain the support functions. 
Instead, we choose to minimize the functional
\begin{eqnarray}\label{eq:trace}
 \min_{\phi_\alpha}\sum_\alpha\Braket{\phi_\alpha|\mathcal{H}_\alpha|\phi_\alpha}\;,
\end{eqnarray}
% where the ``trace'' of the support function dependent Hamiltonian is employed.
while applying both orthogonality and localization constraints on the support functions, as discussed in Section~\ref{sec:imposingloc}.

Apart from the improved localization, the use of the confining potential has yet another advantage. The band structure
energy (Eq.~\eqref{eq:band_structure_energy}) is invariant under unitary transformations among the support functions if
there are no localization constraints. This corresponds to some zero eigenvalues in the Hessian characterizing the
optimization of the support functions. 
The introduction of a localization constraint violates this invariance
and leads to small but non-zero eigenvalues. 
The condition number, defined as the ratio of the largest and smallest
(nonzero) eigenvalue of the Hessian, can thus become very large, potentially turning the optimization into a strongly
ill-conditioned problem.
On the other hand, if the unitary invariance is heavily violated in Eq.~\eqref{eq:trace} by the introduction of a 
strong localization potential, 
the small eigenvalues grow and the condition number improves as a consequence.

After a few iterations of the outer loop, the support functions are sufficiently localized to continue the optimization
without a confining potential, i.e.\ by minimizing the band structure energy. This procedure will lead to highly accurate support functions 
while still preserving locality.
As an alternative it is also possible to define a so-called ``hybrid mode'' which combines the two categories 
of support function optimization and thus provides a smoother transition between the two. In this case the target function is given by
\begin{equation}\label{eq:hybrid}
 \Omega^{hy} = \sum_\alpha K^{\alpha\alpha}\braket{\phi_\alpha|\mathcal{H}_\alpha|\phi_\alpha} + \sum_{\beta\neq\alpha} K^{\alpha\beta}\braket{\phi_\alpha|\mathcal{H}_{KS}|\phi_\beta}.
\end{equation}
In the beginning a strong confinement is used, making this expression similar to the functional of Eq.~\eqref{eq:trace}; 
however the confining
 potential is reduced throughout the calculation so that towards the end the strength of the confinement is negligible and
Eq.~\eqref{eq:hybrid} reverts to the full energy expression.
A prescription for reducing the confinement is presented in Appendix~\ref{app:hybrid_mode}.

\subsubsection{Orthogonalization}
The Lagrange multiplier formalism conserves the orthogonormality of the adaptively-contracted basis only to 
first order. An additional explicit orthogonalization has to be performed after each update of the adaptively-contracted basis 
to restore exact orthogonality. This is done using the L\"owdin procedure.
The calculation of $\mathbf{S}^{-1/2}$, which is required in this context, can pose a bottleneck.
However, as our basis functions are close to orthonormality, the exact
calculation can safely be replaced by a first order Taylor approximation. Numerical tests have shown 
that the error of this approximation is of the same order of magnitude as the inevitable deviation from exact orthonormality 
which is inherent to our set of support functions due to the strictly enforced locality;
consequently the slight non-orthonormality of the adaptively-contracted basis does not significantly increase by the use of the Taylor approximation.

It is important to note that the support functions will only be nearly orthogonal rather than exactly orthogonal, 
as exact orthogonality is in general not possible for functions exhibiting compact support in a discretized space. 
 This near-orthogonality is in contrast with most other 
minimal basis implementations which use fully non-orthogonal support functions~\cite{0953-8984-22-7-074207,skylaris:084119}. 
The asymptotic decay behavior of the orthogonal and non-orthogonal support functions is identical.
However the prefactor differs and leads to a better localization of the non-orthogonal functions~\cite{PhysRevLett.21.13}.
However, in practice, we have found that the introduction of 
the orthogonality constraint does not significantly increase the size required for the localization regions, provided 
that a sufficiently strong confining potential is applied to localize the support functions 
at the start of the calculation.

In order to counteract the small deviations from 
idempotency caused by the changing overlap matrix, we purify the density kernel during the support function optimization either by
directly orthonormalizing the expansion coefficients of the  KS orbitals or by using the McWeeny purification transformation.
% Towards the end of the calculation we can even stop the explicit orthonormalization step in order to converge the final energy to an even higher
%precision, while keeping the orthogonality constraint in the gradient, to ensure the basis functions do not stray too far from
%near-orthogonality.

\subsubsection{Gradient and preconditioning}\label{sec:getlocbasis}
The optimization is done via a direct minimization scheme or with direct
inversion of the iterative subspace (DIIS)~\cite{pulay1980convergence}, both combined with an efficient preconditioning scheme.
The derivation of the gradient of the target function $\Omega$ with respect to the support functions
 involves some subtleties for the trace and hybrid modes since in these cases the Hamiltonian depends
 explicitly on the support
function, leading to an asymmetry of the Lagrange multiplier matrix
 $\Lambda_{\alpha\beta}=\Braket{\phi_\alpha |\sum_\gamma S_{\beta\gamma}\frac{\delta \Omega}{\delta
\Bra{\phi_\gamma}}}$. In order to correctly derive the gradient expression we follow the same guidelines as 
Ref.~\onlinecite{goedecker1997critical}; assuming nearly orthogonal orbitals the final result is given by
\begin{eqnarray}
 \Ket{g_\alpha} = \sum_\beta S_{\alpha\beta}\frac{\delta \Omega}{\delta \Bra{\phi_\beta}} -
\frac{1}{2}\sum_{\beta}\left[\mathbf{S}^{-1}\left(\mathbf{\Lambda}+\mathbf{\Lambda}^T\right)\right]_{\alpha\beta}\Ket{\phi_\beta}.
\label{eq:gradient}
\end{eqnarray}
This is a generalization of the ordinary expression and thus also valid if the
Hamiltonian does not explicitly depend on the support function, i.e.\ for the energy mode. 
As discussed in Section~\ref{sec:imposingloc} the gradient is suitably
localized once derived, i.e.\ $|g_\alpha\rangle \leftarrow \mathcal L^{(\alpha)} | g_\alpha \rangle$.

To precondition the gradient $\Ket{g_\alpha}$ we use the standard kinetic preconditioning
scheme~\cite{genovese:014109}.  To ensure that the preconditioning does not negatively impact the localization of the gradient, we have found that it is important to
add an extra term to account for the confining potential if present.  In this case, the expression
becomes
\begin{eqnarray}
 (-\frac{1}{2}\nabla^2+a_\alpha(\mathbf{r}-\mathbf{R}_\alpha)^4-\varepsilon_\alpha)\Ket{g_\alpha^{prec}}=-\Ket{g_\alpha},
 \label{eq: precond}
\end{eqnarray}
where $\varepsilon_\alpha$ is an approximate value of $\Lambda_{\alpha\alpha}$.  This also has the effect of improving the convergence.
The inclusion of the confining potential adds only a small overhead as it can be
evaluated via convolutions in the same manner as the kinetic energy. 
Furthermore, the preconditioning equations do not need to be solved with high accuracy, only approximately.

\subsection{Density kernel optimization}\label{sec:kernel}

For the optimization of the density kernel we have implemented three schemes: diagonalization, 
direct minimization and the
Fermi operator expansion method (FOE)~\cite{goedecker1994efficient,goedecker1995tight-binding}. Once the kernel has been
updated, we recalculate the charge density via Eq~\eqref{eq:density}; the new
density is then used to update the potential, with an optional step wherein the density is mixed with
 the previous one in order
to improve convergence. This procedure is repeated until the kernel is converged;
in practice, we consider this convergence to have been reached once the mean 
difference of the density of
two consecutive iterations is below a given threshold, i.e.\ $\Delta\rho < c$.

The direct diagonalization method consists of finding the solution of the generalized eigenproblem for a given
Hamiltonian and overlap matrix. Its implementation therefore relies straightforwardly on linear
algebra solvers and will not be detailed here.

In the direct minimization approach the band-structure energy is minimized subject to the
 orthogonality of the support functions. To this end, we express the gradient of the Kohn-Sham orbitals,
$\ket{g_i}=\mathcal{H}_{KS}\ket{\varPsi_i}-\sum_j\Lambda_{ij}\ket{\varPsi_j}$, in terms of the support functions, i.e.\
$\ket{g_i}=\sum_\alpha d_i^\alpha \ket{\phi_\alpha}$. The $d_i^\alpha$ are obtained
by solving
\begin{equation}
 \begin{aligned}
   \sum_\alpha S_{\beta\alpha}d_i^\alpha &= \sum_\alpha H_{\beta\alpha} c_i^\alpha-\sum_j\sum_\alpha S_{\beta\alpha}c_j^\alpha \Lambda_{ji}, \\
    \Lambda_{ji} &= \sum_{\gamma,\delta}c_j^{\gamma *} c_i^\delta H_{\gamma\delta} .
 \end{aligned}
 \label{eq:directmin}
\end{equation}
The coefficients are optimized using this gradient via steepest descents or DIIS. Once the gradient has converged to the required 
threshold, the density kernel
is calculated from the coefficients and occupancies. 

In the Fermi operator expansion method, the density matrix may be defined as a function of the Hamiltonian as
$F=f(\mathcal{H})$, where $f$ is the Fermi function. In terms of the support functions, this corresponds to an
expression for 
the density kernel in terms
of the Hamiltonian matrix, i.e.\ $\mathbf{K}=f(\mathbf{H})$. 
The central idea of the FOE~\cite{goedecker1994efficient,goedecker1995tight-binding,RevModPhys.71.1085} is to find an
expression for $f(\mathbf{H})$ which can be efficiently evaluated numerically.
One particularly simple possibility is a polynomial
expansion; for numerical stability we use Chebyshev polynomials~\cite{press2007numerical}.
As will be shown in detail in Appendix~\ref{AppFOE}, 
the density kernel can be constructed using only matrix vector multiplications thanks to the recursion formulae
for the Chebyshev matrix polynomials.

\subsubsection{Suitability of the methods}

All three methods for calculating the density kernel (direct diagonalization, direct minimization and FOE) yield the same final result, thus the main differences 
lie in their performance, where one of the most important points is the performance of the linear algebra.
Due to the localized nature of the support functions, the overlap, Hamiltonian and density kernel matrices 
are in general sparse, with the level and pattern of sparsity depending on the localization radii of the support
functions and the dimensionality of the system in question.  We can take advantage of this sparsity by 
storing and using these matrices in compressed form and indeed this is necessary to achieve a fully
linear scaling algorithm. 

For diagonalization, exploiting the sparsity is very hard due to 
the lack of efficient parallel solvers for sparse matrices. The method therefore performs badly for large systems
due to its cubic scaling, but thanks to the small prefactor it can still be useful for smaller systems 
consisting of a few hundred atoms.

For direct minimization the situation is better, 
since both the solution of the linear system of
Eq.~\eqref{eq:directmin} and the orthonormalization can be approximated using Taylor expansions for
$\mathbf{S}^{-1}$ and $\mathbf{S}^{-1/2}$, respectively.
It can also be easily parallelized, so that the cubic
scaling terms only become problematic for systems containing more than a few thousand atoms, as demonstrated in
section~\ref{sec:scaling} (see Fig.~\ref{fig:alkane_scaling}). 
Indeed, for moderate system sizes, the extra overhead associated with the manipulation of sparse matrices makes dense matrix
algebra cheaper, and so many physically interesting systems are already considerably accelerated. 

Even though the direct minimization method does not scale linearly in its current implementation, 
there are some situations where its use is advantageous.
For example, the unoccupied states are generally not well represented by the adaptively-contracted basis set following the optimization procedure, 
and for cases where we have only the density kernel and not the coefficients, it becomes necessary
to use another approach to calculate them, 
such as optimizing a second set of minimal basis functions~\cite{PhysRevB.84.165131},
which can be expensive. 
However, as with the diagonalization
approach, the ability to work directly with the coefficients
makes it possible to optimize the support functions
 and coefficients to accurately represent
a few states above the Fermi level at the same time as the occupied states, without significantly impacting the cost.

%It is possible to include a few additional states above the Fermi level in both the support function and coefficient optimization procedures but fixing the occupations of those states to be zero in the charge density calculation, so that the system remains neutral but the support functions are able to accurately represent those extra states.

The FOE approach leads to linear scaling if the sparsity of the Hamiltonian is exploited and if the build 
up of the density matrix is done within localization regions. These localization regions do not need to be identical 
to the localization regions employed for the calculation of the adaptively-contracted basis set -- in general they
will actually be larger.  The final result turns out to be relatively
insensitive to the choice of size for the density matrix localization regions, above a sensible minimum value.
% For medium sized systems the computational savings 
% obtained by the introduction of these extra localization regions are not significant and doing the calculation without 
% them eliminates the need for adjusting them to an appropriate size. 
By exploiting this sparsity, 
we can access system sizes of around ten thousand atoms with a moderate use of parallel resources.

\subsection{Parallelization}
Like standard BigDFT we have a multi-level MPI/OpenMP parallelization scheme~\cite{Genovese2011149}.
The details of the MPI parallelization are presented 
in Appendix~\ref{app:parallelization}.
In Fig.~\ref{fig:speedup_times_withnprocs_withinset_paper} we show the effective
speedup as a function of the number of cores for a large water droplet, 
keeping the number of OpenMP threads at four. 
We measured the parallelization up to 3840 cores, by taking as a reference a 160 core run. As can be seen, the effective 
speedup reaches about 92\% of the ideal value at 480 cores, decreasing to 62\% for 3840 cores.  
Fitting the data to Amdahl's law~\cite{amdahl2007validity} shows that at least 97\% of the code has 
been parallelized; the true value is higher as the times shown include the communications and are relative to 160 cores rather than a serial run. 
Fig.~\ref{fig:speedup_times_withnprocs_withinset_paper} also shows a breakdown of the total calculation 
time into different categories, where we see that the communications start to become limiting 
for the highest number of cores, demonstrating that this is at the upper bound
which is appropriate for this system size.  For a larger system, this limit will of course be higher.

\begin{figure}
 \centering
 \includegraphics[scale=0.35]{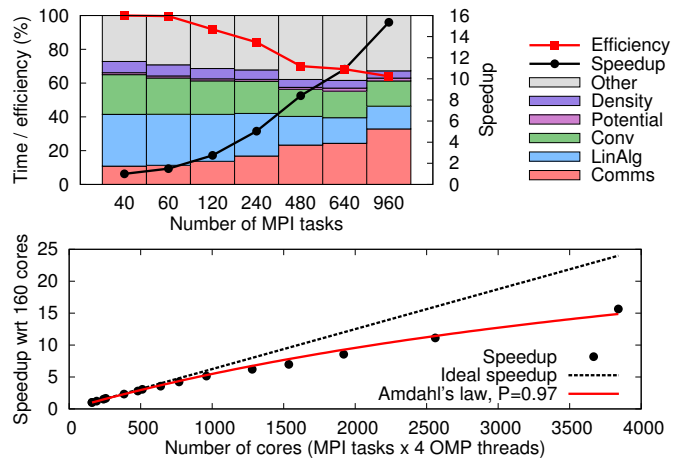}
 \caption{Parallel scaling for a water droplet containing 960 atoms for between 160 and 3840 cores, with the effective and ideal speedup with respect to 160 cores, including a fit of Amdahl's law [below], and a breakdown of the calculation time for selected number of processors [above]. The categories presented are for communications, linear algebra, convolutions, calculation of the potential and density, and miscellaneous.}
 \label{fig:speedup_times_withnprocs_withinset_paper}
\end{figure}

\section{Calculation of ionic forces}\label{pulaysec}

In a self-consistent KS calculation (i.e.\ when the charge density is derived from the numerical set of wavefunctions),
the forces acting on atom $a$ are given by the negative gradient of the 
band structure energy with respect to the atomic positions $\mathbf R_a$.
The Hellmann-Feynman force, given by the expression
\begin{equation}
 \mathbf{F}_a^{(HF)} = -\sum_i f_i \Braket{\varPsi_i|\frac{\partial \mathcal{H}}{\partial\mathbf{R}_a}|\varPsi_i},
\end{equation}
involves only the functional derivative of the Hamiltonian operator.
This term is evaluated numerically in the computational setup used to express the ground state energy.
As explained in more detail in Appendix~\ref{app:Neglect for the traditional cubic version}, with 
the cubic version of BigDFT, only the Hellmann-Feynman term contributes to the forces, as the remaining part tends to zero 
in the limit of small grid spacings. 

However, when the KS orbitals are expressed in terms of the support functions, there is an additional contribution which is not captured by the computational setup. 
As demonstrated in Appendix~\ref{app:The case of the minimal basis setup}, it is given by
\begin{equation}
\mathbf F^{(P)}_a=
-2 \sum _{\alpha, \beta}\mathrm{Re}\left( K^{\alpha \beta}\Braket{\chi_\beta |\frac{\partial \mathcal
L^{(\alpha)}}{\partial \mathbf R_a} |\phi_\alpha}\right),
\end{equation}
where
\begin{equation}
 \Ket{\chi_\alpha} = \mathcal{H}_{KS} \Ket{\phi_\alpha} - \sum_j \sum_{\rho,\sigma} c_j^{\rho} \varepsilon_j c_j^{\sigma *} S_{\sigma \alpha} \Ket{\phi_\rho}
  %\label{eq:support_function_residue}
\end{equation}
is the residual vector of the support function $\Ket{\phi_\alpha}$, which is related to the support function gradient $\Ket{g^\alpha}$ (see Eq.~\eqref{unlocalizedgradient}).
This term can be considered as the equivalent of a Pulay contribution to the ionic forces, arising from the explicit dependence of the localization operators on the atomic positions.

The vector $\frac{\partial \mathcal L^{(\alpha)}}{\partial \mathbf R_a} \Ket{\phi_\alpha}$ only depends on the value of the support function
 on the borders of the localization regions (Eq.~\eqref{eq:locprojderiv}).
Therefore, if the scalar product between the residues $\Ket{\chi_\beta}$ and the values of the support functions $\Ket{\phi_\alpha}$ at the boundaries 
of their localization regions is smaller than the norm of the residue itself (quantifying the accuracy of the results), the Pulay term can be safely neglected.  
%This of course happens when the 

As mentioned in Section~\ref{sec:supportfunc}, the Laplacian operator in the wavelet basis causes
the kinetic energy to be high if the values at the edges are non-negligible.  Such a situation is therefore 
penalized by the energy minimization and so the values at the borders are guaranteed to remain low. 
Indeed, we have seen excellent agreement between the Hellmann-Feynman term only and the forces
calculated using standard BigDFT, as will be demonstrated in sections~\ref{sec:results} and~\ref{sec:geopt}.

The Hellmann-Feynman force is thus the only relevant term even in the adaptively-contracted basis approach and is given by
\begin{eqnarray}
 \mathbf{F}_a^{(HF)} &=&  -\sum_{\alpha,\beta} K^{\alpha\beta} \Braket{\phi_\alpha|\frac{\partial
\mathcal{H}}{\partial\mathbf{R}_a}|\phi_\beta}.
\end{eqnarray}
It is identical to the implementation in standard BigDFT~\cite{genovese:014109} and so the different terms are 
not repeated here. The only difference is that instead of applying the operator to the wavefunctions, 
we now apply it to all overlapping support functions. This can be done efficiently since each support function
 overlaps with only a few neighbors.

\section{Accuracy}
\label{sec:results}

\begin{figure*}
\begin{center}
\subfigure[Cinchonidine ($\mathrm{C}_{19}\mathrm{H}_{22}\mathrm{N}_2\mathrm{O}$), $R_{cut}=$5.29\AA]{\label{fig:cinchonidine}\includegraphics[scale=0.06]{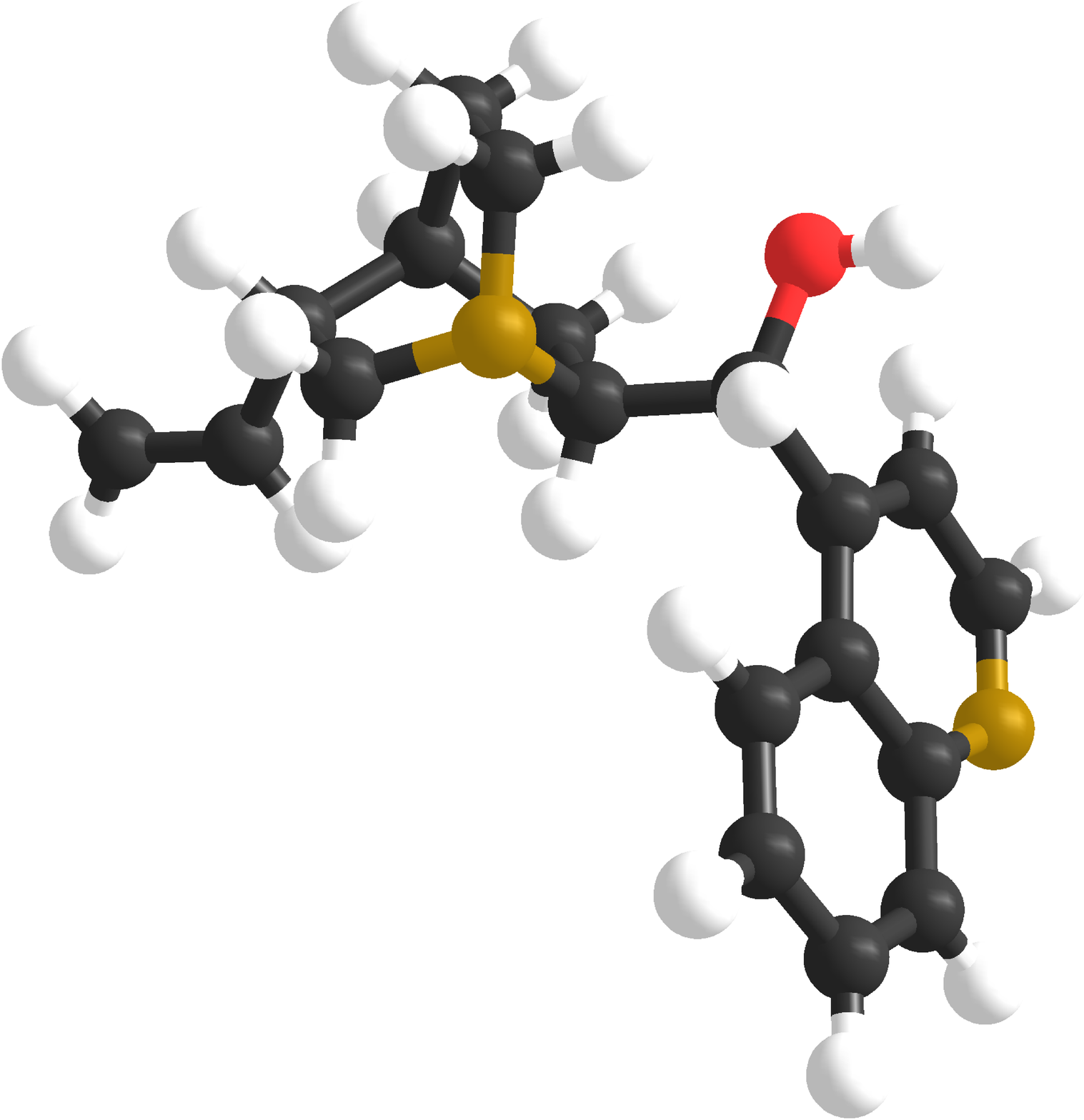}}
\hspace{15pt}
\subfigure[Boron cluster~\cite{PhysRevLett.106.225502}, $R_{cut}=$5.82\AA]{\label{fig:boron}\includegraphics[scale=0.07]{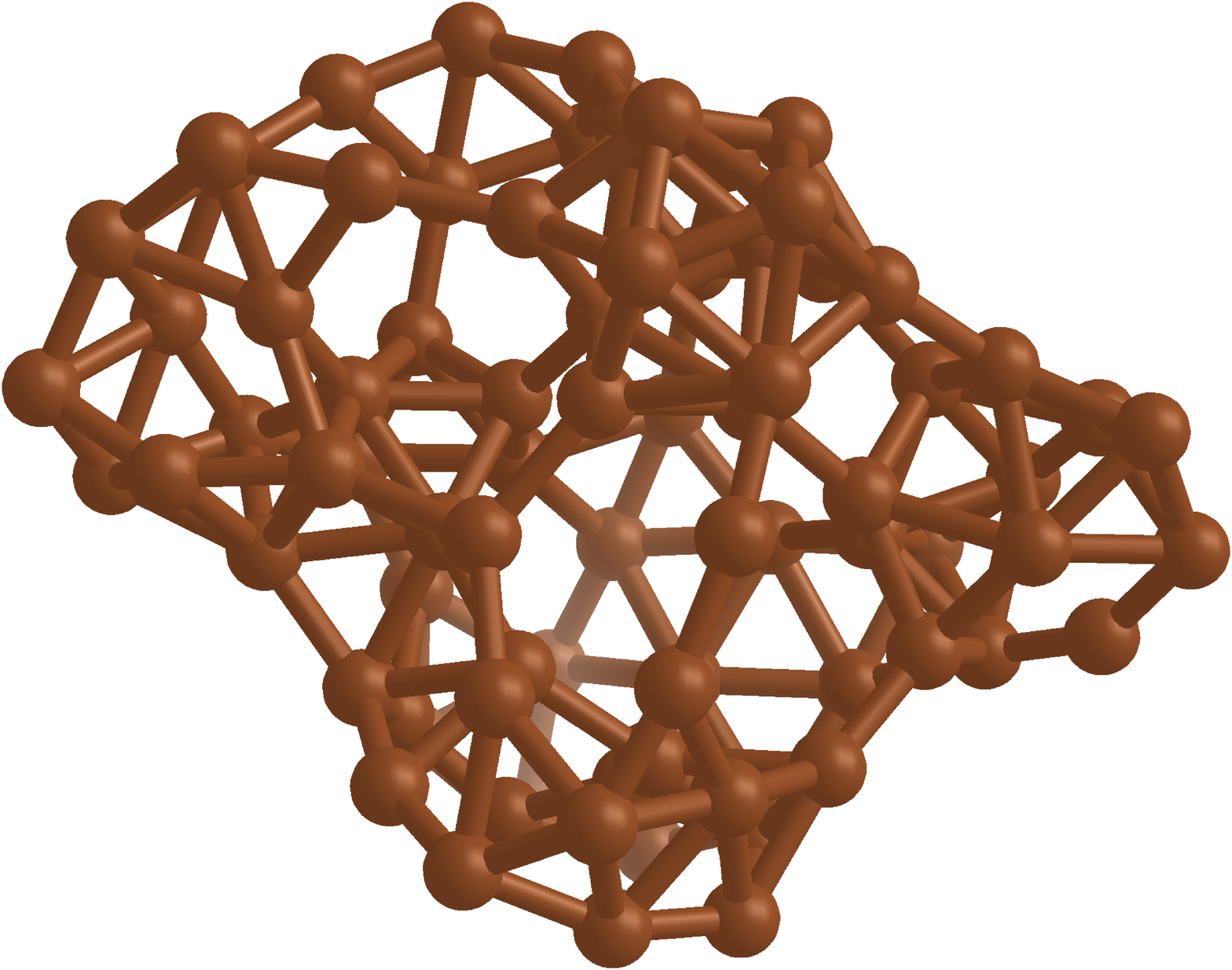}}
\hspace{15pt}
\subfigure[Alkane, $R_{cut}=$5.29\AA]{\label{fig:alkane}\includegraphics[scale=0.06]{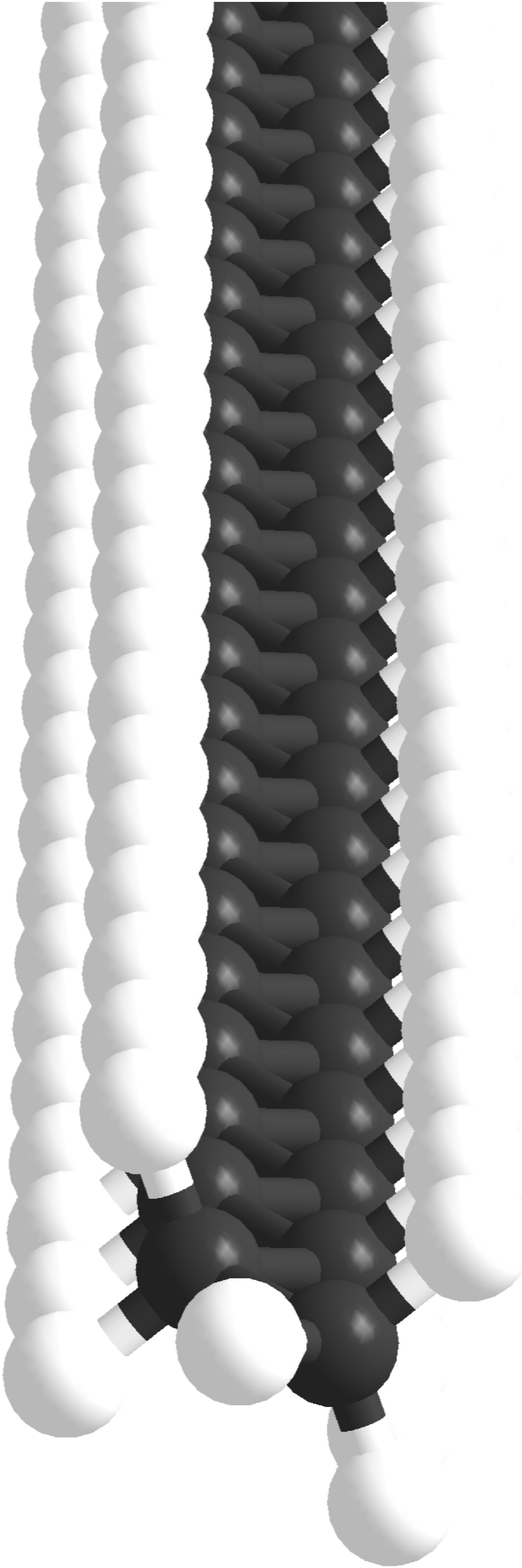}} 
\hspace{15pt}
\subfigure[Water droplet, $R_{cut}=$5.29\AA]{\label{fig:water}\includegraphics[scale=0.06]{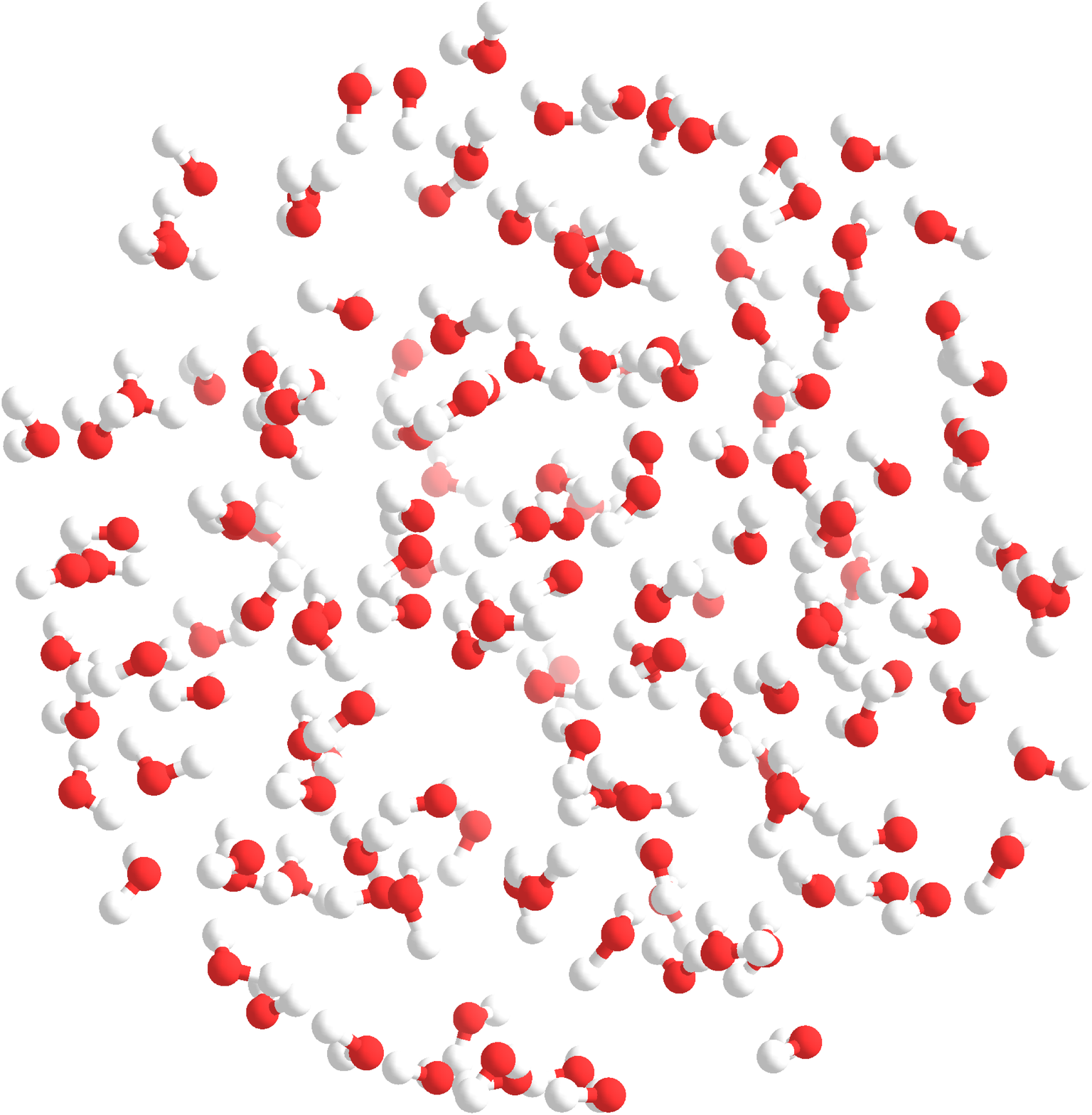}}
\subfigure[$\text{C}_{60}$]{\label{fig:c60}\includegraphics[scale=0.06]{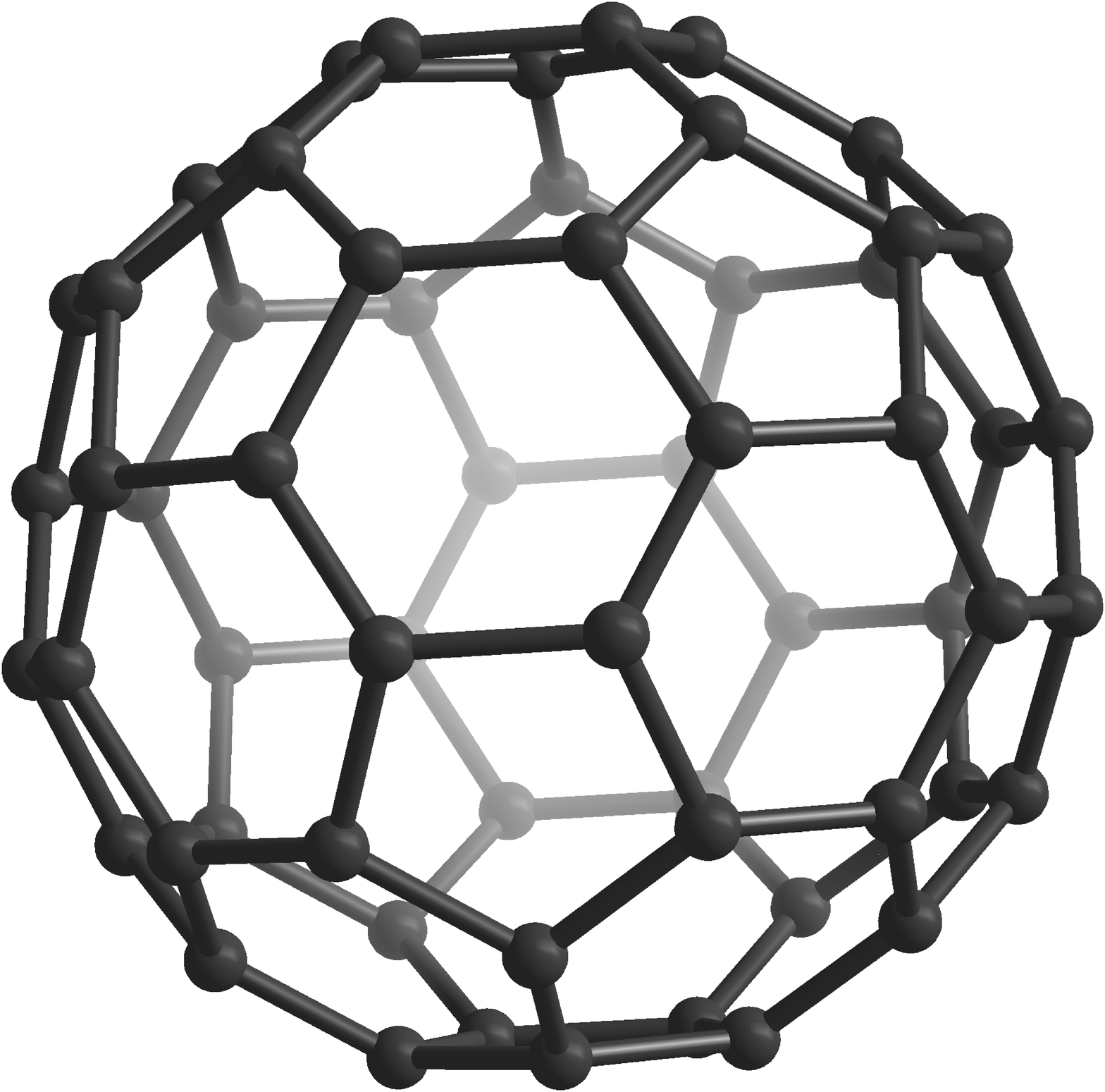}}
\hspace{15pt}
\subfigure[Silicon cluster, $R_{cut}=$4.76\AA]{\label{fig:silicon-cluster}\includegraphics[scale=0.06]{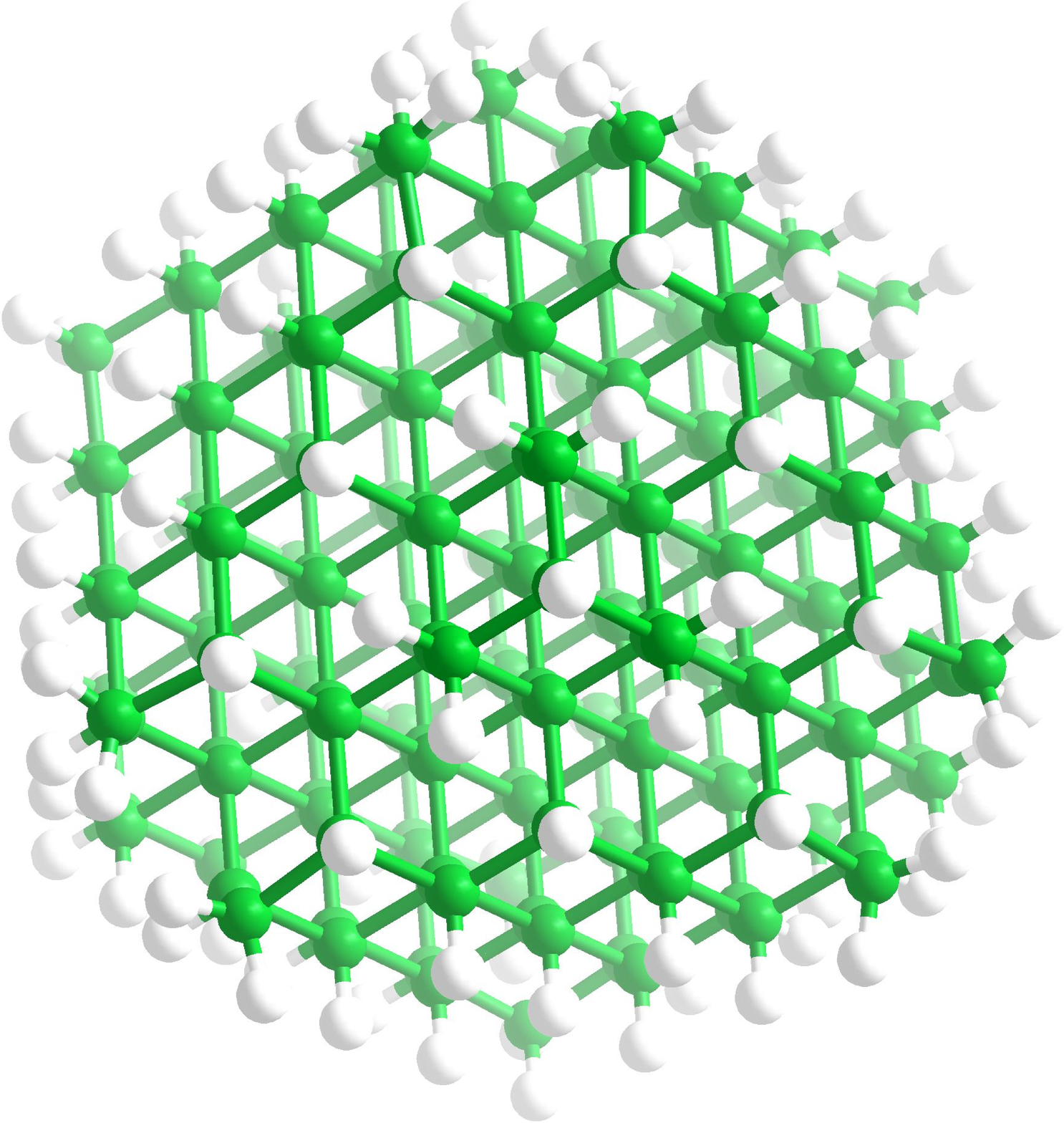}}
\hspace{15pt}
\subfigure[SiC~\cite{PhysRevB.82.035431}, $R_{cut}=$5.82\AA]{\label{fig:sic}\includegraphics[scale=0.06]{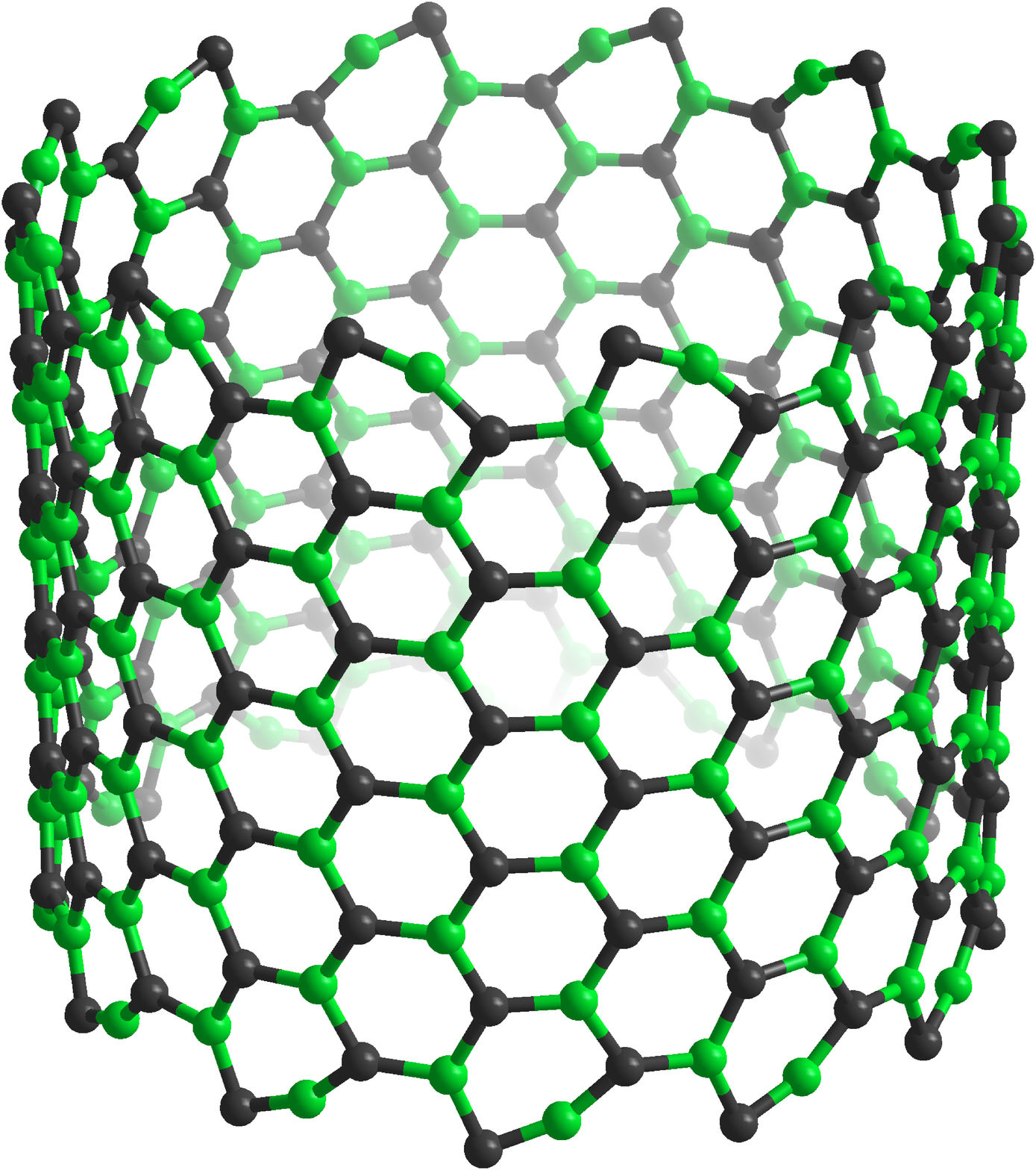}}
\hspace{15pt}
\subfigure[Ladder polythiophene, $R_{cut}=$5.29\AA]{\label{fig:lpt}\includegraphics[scale=0.06]{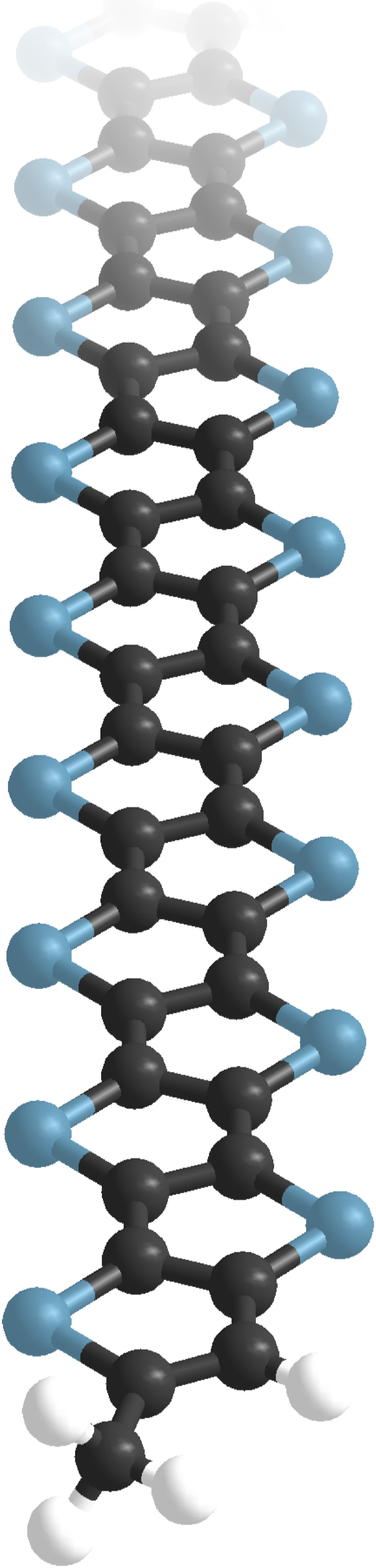}}
\caption{The different systems studied, where gray denotes carbon, white hydrogen, red oxygen, gold nitrogen, bronze
boron, green silicon and blue sulfur.  The values used for the support function localization radii are also given.}\label{fig:accuracies}

\end{center}
\end{figure*}

We have applied our minimal basis approach to a number of systems, depicted in Fig.~\ref{fig:accuracies}, in order to demonstrate both its accuracy 
and its applicability.
% These include tests of the accuracy of both energies and forces, including their convergence with respect to the localization
% radius.  We have also tested the performance of the method, including scaling with respect to system size, the convergence
% behaviour for different system sizes, the performance 
% of a geometry optimization and calculations on a charged system.  
All calculations have been 
done using the local density approximation (LDA) exchange-correlation functional~\cite{PhysRevLett.45.566} and HGH
pseudopotentials~\cite{PhysRevB.58.3641}. However it is worth noting that the sole usage of LDA does not imply a general restriction and 
other functionals can be used as well; as an example, a PBE~\cite{Perdew1996} calculation is presented for one system. In addition, we have used free boundary conditions, avoiding the need for the
supercell approximation.  
The values of the wavelet basis parameters for the different systems, as well as the localization radii, were selected 
in order to achieve accuracies better than \unit[1]{meV/atom}. 
This corresponded to values between \unit[0.13]{\AA} and \unit[0.20]{\AA} for $h$, 5.0 and 7.0
for $\lambda^c$ and 7.0 and 8.0 for $\lambda^f$.
% and they are summarized in Tab.~\ref{table:hgrids}. 
Unless otherwise stated, we have used the direct 
minimization scheme for the density kernel optimization.
For hydrogen atoms one basis function was used per atom, 
whereas for all other
elements four basis functions were used per atom, except where otherwise stated.

%\subsection{Accuracy: energy and forces} \label{sec:results_accuracy}

\subsection{Benchmark systems}\label{sec:results_accuracy}
\begin{figure}
\begin{center}
\includegraphics[scale=0.35]{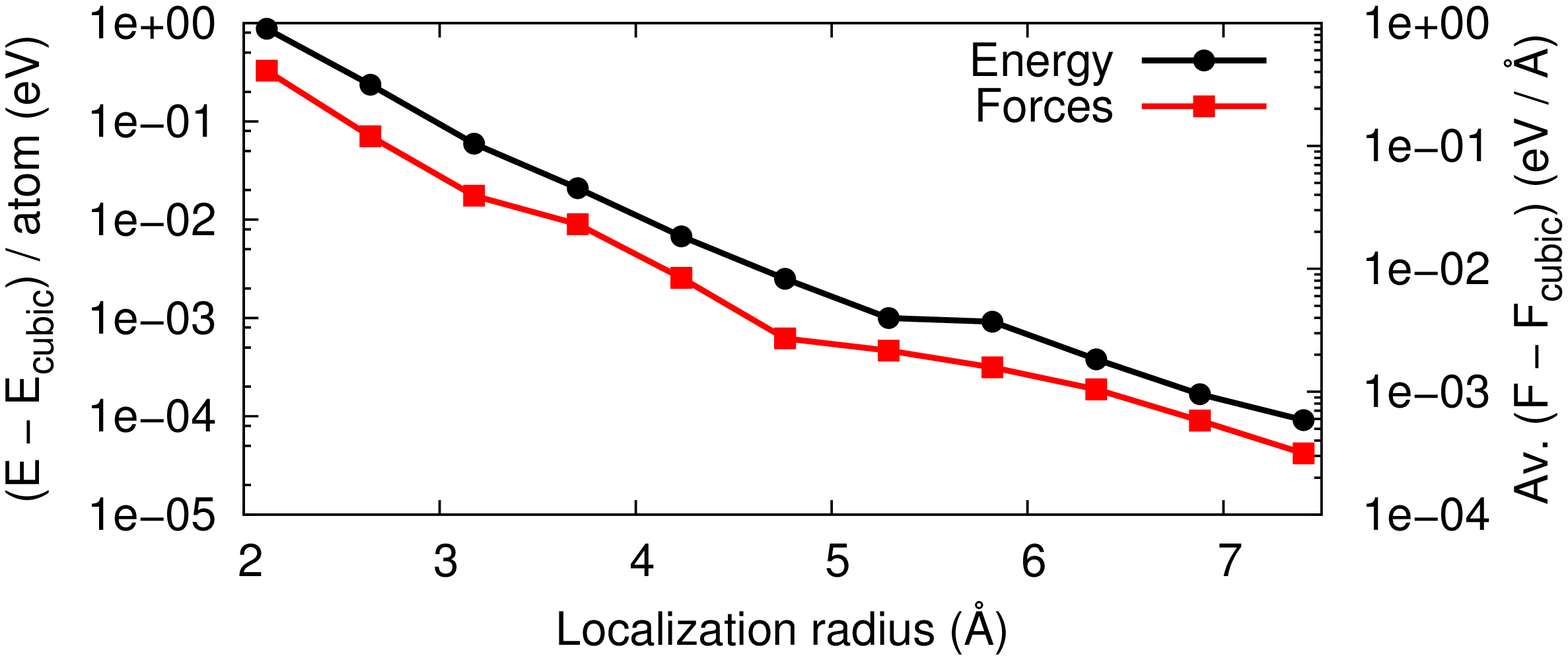}
\caption{Convergence of the total energy and forces with respect to the localization radius as 
compared to the cubic values for a molecule of $\text{C}_{60}$. For large localization regions 
the accuracy of the forces is sufficient for high accuracy geometry optimizations.}
\label{fig:locrad_conv}
\end{center}
\end{figure}

We demonstrate excellent agreement with the traditional
cubic scaling method for both energy and forces -- of the order of \unit[1]{meV/atom} 
for the energy and a few~\unit[]{meV/\AA} for the forces, as shown in 
Tab.~\ref{table:accuracies}. 
We also demonstrate systematic convergence of the
total energy and forces with respect to localization radius for 
a molecule of $\text{C}_{60}$, where
the largest localization regions are close to the total system size, as depicted
in Fig.~\ref{fig:locrad_conv}.
For all these systems
the level of accuracy achieved for the forces using the
Hellmann-Feynman term only is
of the same order as that of the cubic code.

% \begin{table}
% \centering
% \begin{tabular}{lcccc}
% \hline\hline
% & $h$  (\AA) & $\lambda^c$ & $\lambda^f$ & $R_{cut}$ (\AA)\\
% \hline
% Cinchonidine & 0.159 & 5.0 & 7.0 & 5.29 \\
% $B$ cluster & 0.201 & 5.0 & 7.0 & 5.82\\
% Alkanes & 0.185 & 6.0 & 8.0  & 5.29\\
% Water & 0.132 & 7.0 & 7.0 & 5.29 \\
% $\text{C}_{60}$ & 0.185 & 6.0 & 8.0 & Fig.~\ref{fig:locrad_conv}\\
% Si cluster & 0.185 & 5.0 & 7.0 & 4.76 \\
% SiC cage & 0.185 & 5.0 & 7.0 & 5.82 \\
% LPT & 0.185 & 6.0 & 8.0 & 5.29\\
% \hline\hline
% \end{tabular}
% \caption{Values of the parameters for the wavelet basis set and localization radii used
% for defining $\mathcal L^{(\alpha)}$ projectors for support functions. The systems are depicted in Fig.~\ref{fig:accuracies}}\label{table:hgrids}
% \end{table}

\begin{table*}
\centering
\begin{tabular}{lc|cc|c||cc|c}
\hline\hline
\multirow{2}{*}{ } & \multirow{2}{*}{Num. atoms} & \multicolumn{3}{c||}{Energy (eV)} & \multicolumn{3}{c}{Forces (eV/\AA)}\\[1.0ex]
& & Min. Basis & Cubic & (Min. - Cub.)/atom & Min. Basis & Cubic & Av. (Min. - Cub.) \\
\hline
Cinchonidine LDA & 44 & $-4.273\cdot 10^{3}$ & $-4.273\cdot 10^{3}$ & $2.0\cdot 10^{-3}$ & $2.734\cdot 10^{-2}$ & $2.772\cdot 10^{-2}$ & $4.3\cdot 10^{-3}$ \\[0.5ex]
Cinchonidine PBE & 44 & $-4.274\cdot 10^{3}$ & $-4.274\cdot 10^{3}$ & $3.0\cdot 10^{-4}$ & $2.581\cdot 10^{-2}$ & $2.610\cdot 10^{-2}$ & $2.2\cdot 10^{-3}$ \\[0.5ex]
Boron cluster & 80 & $-6.141\cdot 10^{3}$ & $-6.141\cdot 10^{3}$ & {$2.4\cdot 10^{-3}$} & $1.305\cdot 10^{-2}$ & $1.316\cdot 10^{-2}$ & $4.2\cdot 10^{-3}$ \\[1.0ex]
Alkane & 257 & $-1.592\cdot 10^{4}$ & $-1.592\cdot 10^{4}$ & {$3.7\cdot 10^{-4}$} & $3.760\cdot 10^{-2}$ & $3.767\cdot 10^{-2}$ & $1.0\cdot 10^{-3}$ \\[0.5ex]
Water & 450 & $-7.011\cdot 10^{4}$ & $-7.012\cdot 10^{4}$ & {$1.7\cdot 10^{-3}$} & $3.730\cdot 10^{-2}$ & $3.730\cdot 10^{-2}$ & $2.5\cdot 10^{-3}$ \\[0.5ex]
\hline\hline
\end{tabular}
\caption{Energy and force differences between the minimal basis approach and the standard cubic
 version of the code.  For the forces, the root mean squared force is given for each 
approach, as well as the average difference between each component.}  \label{table:accuracies}
\end{table*}

\subsection{Silicon defect energy}
In order to demonstrate the accuracy of our method for a practical application we calculated the energy of 
a vacancy defect in a hydrogen terminated silicon cluster containing 291 atoms, shown in Fig.~\ref{fig:silicon-cluster}.
As shown in Tab.~\ref{tab:silicon-clusters}, the difference in defect energy
between the cubic reference calculation and the linear version is \unit[129]{meV} using 4 support 
functions per Si atom and 1 support function per
hydrogen atom. Even more accurate results can be achieved by
increasing the number of support functions per Si atom to 9, which reduces the error to \unit[12]{meV}. 
Increasing the localization 
radii does not further improve the accuracy as the result is already within the noise level.
To achieve these results, support functions were optimized using the hybrid mode and the density kernel was
optimized using the FOE approach with a cutoff of \unit[7.94]{\AA} for the kernel construction. 

%\end{figure}
 \begin{table}
  \centering
  \begin{tabular}{l r r r c r r r c r}
   \hline\hline
          & \multicolumn{1}{c}{pristine} & \multicolumn{1}{c}{vacancy} & \multicolumn{1}{c}{$\Delta$} & \multicolumn{1}{c}{$\Delta$-$\Delta_{\text{cubic}}$} \\
          & \multicolumn{1}{c}{eV} & \multicolumn{1}{c}{eV} & \multicolumn{1}{c}{eV} & \multicolumn{1}{c}{meV} \\
 \hline
    cubic & $-20674.223$ & $-20563.056$ & $111.167$ & -- \\ 
    4/1 & $-20667.556$ & $-20556.518$ & $111.038$ & $129$ \\
    9/1 & $-20672.856$ & $-20561.701$  & $111.155$ &  $12$ \\
   \hline\hline
  \end{tabular}
  \caption{Total energies and energy differences for a silicon cluster (Fig.~\ref{fig:silicon-cluster}) with
 and without a vacancy defect. The first column shows the number 
of support functions per Si and H atom, respectively, and the last column shows the difference in defect energy 
between the cubic and linear versions.}
  \label{tab:silicon-clusters}
 \end{table}

\subsection{Consistency of energies and forces}\label{ap:forces}

Following the discussion in Section~\ref{pulaysec}, we have calculated the average value of the support functions on the borders of their localization regions for various systems and found this to be
at least three orders of magnitude smaller than the norm of the support function residue  
(defined in Eq.~\eqref{eq:KS_residue}).  This is in line with our expectations, as discussed in Section~\ref{pulaysec}, and implies
that the Pulay terms should be negligible compared to the error introduced to the Hellmann-Feynman term due to the localization constraint.
Indeed, this agrees with the calculated forces for the systems presented thus far. 
To further verify that the Pulay term can be neglected and to quantify the different sources of errors,
 we have also checked that the calculated forces are consistent with the energy, i.e.\ that they correspond to
its negative derivative. To this end, initial and final configurations $\mathbf{R}^{(a)}$ and
$\mathbf{R}^{(b)}$ of a given system were chosen, where $\mathbf{R}$ represents the atomic positions. 
Small steps $\Delta \mathbf{R}$ were then taken between $\mathbf{R}^{(a)}$ to $\mathbf{R}^{(b)}$. 
If the forces $\mathbf{F}$ are correctly evaluated we should have
\begin{equation}
 \Delta E = \int_a^b \mathbf{F}(\mathbf{r})\cdot\mathrm{d}\mathbf{r} \approx \sum_{\mu=a}^b \mathbf{F}(\mathbf{R}_\mu)\cdot\Delta\mathbf{R}_\mu,
\end{equation}
where $\mu$ labels the intermediate steps between
 configurations $\mathbf{R}^{(a)}$ and
$\mathbf{R}^{(b)}$.  This approximation can be compared with the exact value obtained by directly
calculating the energy differences, i.e.\ $E(\mathbf{R}^{(b)})-E(\mathbf{R}^{(a)})$. These two values should agree
with each other up to the noise level of the calculation.

To analyze the different terms contributing to the noise in the forces, we use a combination of the hybrid and FOE methods with five progressive setups which
give an estimate of the magnitude of the various error sources:
\begin{enumerate}
 \item Using the cubic scaling scheme where all 
 orbitals can extend over the full simulation cell.
 \item Using the minimal basis approach but without localization constraints or 
 confining potential.
 \item Applying a confining potential but no localization constraints.
 \item Using a finite localization radius of \unit[7.94]{\AA} for the density kernel, but not for the support functions.
 \item Applying in addition strong localization radii of \unit[4.76]{\AA} to the support functions.
\end{enumerate}
This test was done for a 92 atom alkane -- despite the relatively small system size the introduction of 
 finite cutoff radii for the support functions and the density kernel construction has a strong effect 
since for chain-like structures the volume of the localization region is only a small fraction of the total computational volume. 
The results are shown in Tab.~\ref{tab:comparison_forces}, for a step size of $\Delta\mathbf{R}_\mu = \unit[0.003]{\AA}$. As expected, without the application of the localization
constraint, the
 errors for the minimal basis calculations are of the same order of magnitude as the reference cubic calculation.  This
is also the case when 
a finite cutoff radius for the construction of the density kernel is introduced. Once a finite localization is imposed
 on the support functions the discrepancy between the energy difference and the force integral increases by an order of
magnitude, however it remains small, agreeing with our previous observations about the Pulay
forces for large enough localization radii.

\begin{table}[h!]
 \small
 \centering
 \begin{tabular}{ccccc r r r}
  \hline\hline
 \multicolumn{1}{p{15pt}}{\rotatebox[origin=c]{70}{Setup}} & \multicolumn{1}{p{15pt}}{\rotatebox[origin=c]{70}{Sup. func.}} & \multicolumn{1}{p{15pt}}{\rotatebox[origin=c]{70}{Conf.\ pot.}} & \multicolumn{1}{p{15pt}}{\rotatebox[origin=c]{70}{$K^{\alpha\beta}$ cutoff}} & \multicolumn{1}{p{15pt}}{\rotatebox[origin=c]{70}{$\phi_\alpha$ cutoff}} & \multicolumn{1}{c}{$\Delta E$} & \multicolumn{1}{c}{$\int \mathbf{F}(\mathbf{r})\cdot\mathrm{d}\mathbf{r}$} & \multicolumn{1}{c}{diff.} \\
\hline
1 & \ding{55} & \ding{55} & \ding{55} & \ding{55} & 5.666961    & 5.667190     & $-2.3 \cdot 10^{-4}$           \\
2 & \ding{51} & \ding{55} & \ding{55} & \ding{55} & 5.666966    & 5.666999     & $-3.3 \cdot 10^{-5}$           \\
3 & \ding{51} & \ding{51} & \ding{55} & \ding{55} & 5.666958    & 5.667024     & $-6.5 \cdot 10^{-5}$           \\  
4 & \ding{51} & \ding{51} & \ding{51} & \ding{55} & 5.667239    & 5.667024     & $ 2.1 \cdot 10^{-4}$           \\
5 & \ding{51} & \ding{51} & \ding{51} & \ding{51} & 5.669992    & 5.673043     & $-3.1 \cdot 10^{-3}$           \\
  \hline\hline
 \end{tabular}
 \caption{Force calculations for the five setups described in the text, with all quantities given in eV.}
 \label{tab:comparison_forces}
\end{table}

\section{Scaling and crossover point}\label{sec:scaling}

\begin{figure}
\begin{center}
\includegraphics[scale=0.35]{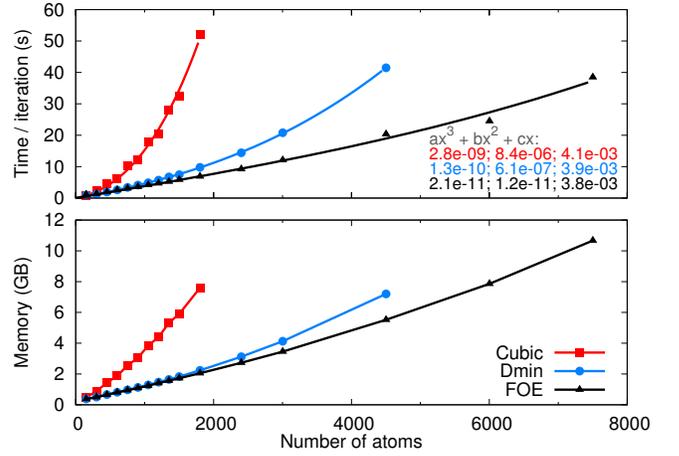}
\caption{Comparison between the time taken per iteration and memory usage for the cubic scaling
and minimal basis approaches using both direct minimization (Dmin) and FOE
for increasing length alkanes, where the time is 
for the wavefunction optimization, neglecting the input guess.
The coefficients are shown for the corresponding cubic polynomials.
A fixed number of 301 MPI tasks and 8 OpenMP threads was used.
}
\label{fig:alkane_scaling}
\end{center}
\end{figure}

We have applied the minimal basis method to alkanes, applying both the direct minimization and FOE approaches for the
kernel optimization.  The time taken per iteration is compared with the traditional cubic-scaling version
 in Fig.~\ref{fig:alkane_scaling}. The number of iterations required to reach convergence 
is similar for the cubic and minimal basis approaches and is approximately constant across system sizes,
so that the total time taken shows similar behaviour.
The results clearly demonstrate the improved scaling of the method, with a crossover point for the total time
at around 150 atoms.  This will of course be system dependent -- the chain like nature of the alkanes makes them a particularly
favorable system for the minimal basis approach.  We also plot cubic polynomials for the 
timing data;  
whilst the cubic scaling approach only has a very small cubic term, both this and the quadratic term are noticeably
reduced for both the FOE and direct minimization approaches.  Indeed, the FOE method
is predominantly linear scaling, compared to direct minimization which has larger quadratic and cubic terms, mainly due to the linear
algebra, as expected.  

The minimal basis approach also gives considerable savings in memory; for the above example the memory requirements for the cubic
 version still prohibit calculations on systems bigger than around 2000 atoms for the chosen number of processors, whereas for the minimal basis method the
memory requirements allow calculations of up to 4000 atoms using direct minimization,
and nearly 8000 atoms with FOE.

\begin{figure}
\begin{center}
\includegraphics[scale=0.35]{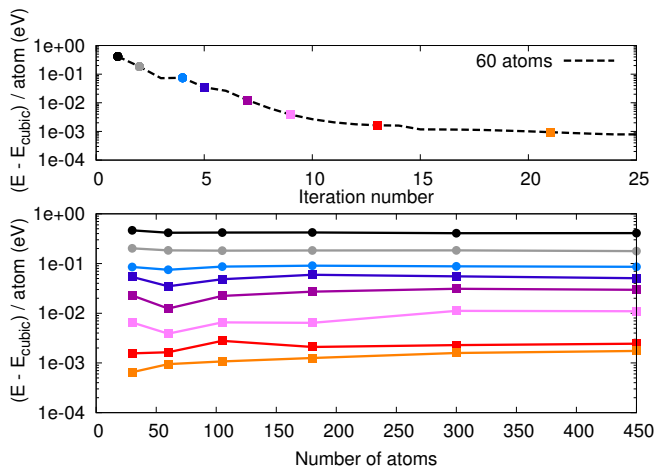}
\caption{Convergence behavior for water droplets, showing the
 convergence with respect to outer loop iteration number for the system containing 60 atoms [above] and the
 energy difference with the converged cubic value after certain numbers of iterations [below].  The number of iterations is
indicated by the color used; circles denote iterations with a confining potential and squares without.}
\label{fig:size_conv}
\end{center}
\end{figure}

To take full advantage of the improvements made to BigDFT, it is not enough for the time
taken per iteration to scale favorably with respect to system size, it is also necessary for the
number of iterations needed to reach convergence not to increase with system size. 
We have demonstrated such behavior for increasing sized randomly generated non-equilibrium water droplets, as shown in
Fig.~\ref{fig:size_conv}.  The number of iterations required to reach a
good level of agreement with the cubic scaling version of the code remains approximately constant, with the 
fluctuations due to the random noise in the bond lengths of the water molecules.  Furthermore, the energy
converges rapidly to a value very close to that obtained with the cubic code, as illustrated by the upper panel.  We have also observed similar
convergence behavior for other systems, including alkanes, as mentioned above.

\section{Flexibility of the minimal basis formalism}
\subsection{Geometry optimization} \label{sec:geopt}
As a further test of the quality of the forces and as a demonstration of the flexibility of the minimal basis formalism, we
 have performed a geometry optimization for a segment of a SiC nanotube containing 288
 atoms, depicted in Fig.~\ref{fig:sic}. Here we can take advantage of the minimal
 basis formalism by reusing the optimized support functions from the previous geometry step as an improved input guess, 
moving them with the atoms using an interpolation scheme to account for atomic displacements which are not multiples of the 
grid spacing $h$.
This has the effect of 
reducing the number of iterations required to converge the support functions for each new geometry.
In fact, for cases 
where the atoms have only moved a small amount, they will hardly need optimizing at all and so substantial
 savings can be made. 
A similar procedure also exists for the cubic version, but the minimal basis approach can profit much more
because of the direct relation between the support function centers and the atomic positions.

\begin{figure}
\begin{center}
\includegraphics[scale=0.35]{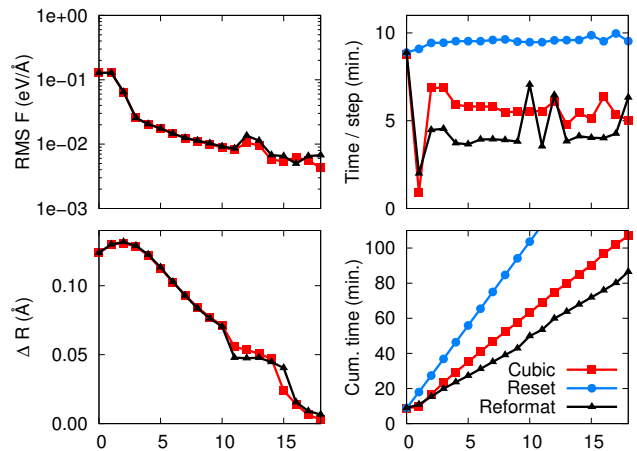}
\caption{Geometry optimization for a segment of a SiC nanotube for the cubic and minimal basis approaches
 where the support functions are regenerated from atomic orbitals at each geometry step (``reset'') and where they are reformatted for
 reuse at the next iteration (``reformat'').  The time taken for each step, cumulative time, force convergence and average distance from the final
 structure optimized using standard BigDFT are plotted for each step of the geometry optimization.}
\label{fig:geopt}
\end{center}
\end{figure}

% This reuse of course necessitates a scheme for reformatting the support functions, as in general the atoms
% will not have moved by an integer number of grid points; this is achieved using an interpolation in real space.  
We compared the convergence behavior and time taken for the minimal basis approach both with and without reusing the support
functions at each geometry step with that for the standard cubic approach, for which the results are shown in Fig.~\ref{fig:geopt}.
It is clear that the Hellmann-Feynman 
forces are sufficiently accurate to optimize the structure to the required level -- in this case forces of below
\unit[$10^{-2}$]{eV/\AA} are readily achieved. For this system size we are below the crossover point, such that 
when the support functions are reset at each geometry step the time taken per step is greater than that for the cubic
approach. However, the reuse of support functions results in a significant reduction in the number of steps
required to fully converge the support functions, and so the total time is less than that required for the
 cubic approach.  This means that the crossover point will be reduced for geometry optimizations or molecular
dynamics calculations, opening up further possibilities for the highly accurate study of dynamics of large systems.

\subsection{Charged systems}

As previously mentioned, the ability to use free boundary conditions 
is essential for charged systems.  This has enabled us to perform calculations of isolated segments of ladder polythiophene (LPT) (Fig.~\ref{fig:lpt}), initially
in a neutral state and then adding a charge of plus or minus two electrons. 
The support functions from the neutral case are also well suited to the charged system
so that only kernel optimizations are required, which can reduce the computational cost by an order of magnitude. 

\begin{table}
\centering
\begin{tabular}{rrcc}
\hline\hline
Q & & {$\Delta E_Q$} & {$|\Delta(E_Q-E_0)|$}\\
\hline
{$0$} & opt. & 176 & --\\
\hline
\multirow{2}{*}{$-2$} & opt. & 147 & 28\\
& unopt. & 292 & 117 \\
\hline
\multirow{2}{*}{$+2$} & opt. & 128 & 47 \\
& unopt. & 200 & 24\\
\hline\hline
\end{tabular}
\caption{Energy differences between the standard cubic
 version and the minimal basis approach for absolute energies ($\Delta E_Q$) and energy differences with the neutral system ($|\Delta(E_Q-E_0)|$) for 63 atom segments of LPT with a net charge $Q$.  Results are shown both for fully optimized support functions (``opt'') and for support functions reused from the neutral calculation (``unopt'').  All results are in meV.}\label{tab:lpt}
\end{table}
In Tab.~\ref{tab:lpt} we compare the agreement between the minimal basis approach and the standard cubic approach for
 a system containing 63 atoms.  We demonstrate an agreement of the order of \unit[100]{meV} for the energy differences for both
 the fully optimized set of support functions and the reuse of the support functions from the neutral system.  For the negatively charged calculations we have also confirmed
 that this level of accuracy is maintained up to 300 atoms, beyond
which size the cost of calculations with the cubic version of BigDFT increases significantly.  

In order to converge the results obtained with the minimal basis to a good level of
accuracy, we used 9 support functions for carbon and sulfur and 1 per hydrogen. 
For charged systems we have found that the direct minimization method is more stable, as it
allows us to update the coefficients in 
smaller steps before updating the 
kernel and therefore density, rather than fully converging them before each update.

We expect such support function reuse to be generally applicable for systems where the addition of a charge only results in a perturbation of the electronic structure. However it may be necessary to optimize a few unoccupied states (using direct minimization or diagonalization)
in order to ensure that the adaptively-contracted basis is sufficiently accurate for negatively charged systems.

\section{Conclusion}

We have presented a self-consistent minimal basis approach within BigDFT which leads to a reduced 
scaling behavior with system size and allows the treatment of larger systems than can be treated with the cubic version;
for very large systems linear scaling is clearly visible. 
The use of a small set of nearly orthogonal adaptively-contracted basis functions which 
are optimized \textit{in situ} in the underlying wavelet basis set gives rise to sparse matrices of relatively small size. 
For the optimization of these so-called support 
functions we use a confining potential which on the one hand helps to keep the support 
functions strictly localized, and on the other hand helps to alleviate the notorious ill-conditioning
which is typical of linear scaling approaches. 

% It was shown that the use of a small set of quasi-orthogonal basis functions is attractive because it can 
% lead to sparser matrices, while the use of a confining potential enables the functions to remain well localized
% and furthermore has the added benefit of improving the
%  conditioning of the algorithm. In principle, instead of a user defined potential, one could use the self-interaction
%  correction potential.                      

%For most applications, like geometry optimization or molecular dynamics, accurate ionic forces are needed. To this end, Pulay corrections become important when localization is imposed. Using convolutions and fast-wavelet transforms, these corrections are evaluated efficiently and almost without any overhead.  

The standard cubic scaling version of BigDFT has been previously demonstrated to give highly accurate results and so we use this as a standard of comparison for our method.
We have demonstrated for a number of different systems excellent agreement
 with the cubic version for both energy and forces.  In particular, we have demonstrated that it is not necessary to include Pulay-like correction terms to the atomic forces, thanks to the nature of the Laplacian operator in the wavelet basis which ensures the support functions remain negligible on the borders of the localization regions.
  In addition, we have shown
consistent convergence behavior across a range of system sizes.
From the viewpoint of scaling with the number of atoms we have demonstrated 
linear scaling for the FOE method 
where the linear algebra has been written to exploit the sparsity of the matrices. 

Finally, we have highlighted some of the advantages of using localized support functions expressed in a wavelet basis set.  These
 include the ability to further accelerate geometry optimizations by reusing the support functions from the previous geometry,
 and the possibility of achieving
a good level of accuracy for a charged calculation by reusing the support functions from a neutral calculation. 

By directly working in the basis of the support functions, we can therefore reduce the number of degrees of freedom needed to express the KS operators for a targeted accuracy.
Aside from reducing the computational overhead, this flexible approach 
paves the way for 
future developments, where the adaptively-contracted basis functions can be reused in other situations, including for example
constrained DFT calculations of large systems. Work is ongoing in this direction.

The authors would like to acknowledge funding from the European project MMM@HPC (RI-261594), the CEA-NANOSCIENCE BigPOL project, the ANR projects SAMSON (ANR-AA08-COSI-015) and NEWCASTLE, and the Swiss CSCS grants s142 and h01.
CPU time and assistance were provided by CSCS, IDRIS, Oak Ridge National Laboratory and Argonne National Laboratory.

\appendix
\section{Prescription for reducing the confinement}
\label{app:hybrid_mode}
To derive a prescription for reducing the confinement, it is assumed that the change in the 
target function between successive iterations of the minimization procedure can be approximated to first order by
\begin{equation}
 \Delta \Omega'_{(n)} = \sum_\alpha \braket{g_{(n)}^\alpha|\Delta\phi_{(n)}^\alpha},
\end{equation}
 where $\ket{\Delta\phi_{(n)}^\alpha}$ is the change in support function between iterations $n$ and $n+1$, i.e.\ $\ket{\Delta\phi_{(n)}^\alpha}=\ket{\phi_{(n+1)}^\alpha}-\ket{\phi_{(n)}^\alpha}$,
and $\ket{g_{(n)}^\alpha}$ is the gradient of the target function with respect to the support function at
iteration $n$. 
Due to the influence of the confinement and the localization regions, the gradient of the support functions and
 thus $\Delta \Omega'_{(n)}$ will not go down to zero. However, the actual change in the target function, $\Delta
\Omega_{(n)}$, will at some point go to zero, meaning that further optimization becomes impossible
for the localization region and confining potential currently used. In this case, the only way to further
  minimize the target function is to decrease the confining potential. Therefore at each step of the 
 minimization the ratio between the actual and estimated decreases in the target function is determined:
\begin{equation}
 \kappa=\frac{\Delta\Omega_{(n)}}{\Delta\Omega'_{(n)}}.
 \label{eq:fraction_to_reduce_the_confinement}
\end{equation}
This value is then used to update the confinement prefactor, $a_\alpha$, at the start of the 
following support function optimization loop, via
\begin{equation}
 a_\alpha^{new}=\kappa a_\alpha^{old}.
 \label{eq:hybrid_mode_prescription_to_reduce_the_confinement}
\end{equation}
If $\kappa$ is of the order of one, this implies there is still some scope for optimizing the support functions using the
 current confining potential and it should not be updated. If, on the other hand, $\kappa$ is much smaller, 
it will hardly be possible to further improve the support functions and so the magnitude of the 
 confining potential should be decreased. In this way one gets a smooth transformation from the hybrid expression to the
energy expression.

\section{Fermi operator expansion} \label{AppFOE}
In the FOE method the density kernel is given as a sum of Chebyshev polynomials.
Since these polynomials are only defined in the interval $[-1,1]$, it is necessary to shift and scale the Hamiltonian 
such that its eigenvalue spectrum lies within this interval. If $\varepsilon_{min}$ and
$\varepsilon_{max}$ are the smallest and largest eigenvalues that would result from diagonalizing the Hamiltonian matrix
according to $\mathbf{H}\mathbf{c}_i=\varepsilon_i\mathbf{S}\mathbf{c}_i$, then the scaled
 Hamiltonian, $\tilde{\mathbf{H}}$,
has to be built using
\begin{equation}
 \tilde{\mathbf{H}} = \sigma(\mathbf{H}-\tau \mathbf{S}),
 \label{eq:scale_and_shift_Hamiltonian}
\end{equation}
with
\begin{equation}
\sigma=\frac{2}{\varepsilon_{max}-\varepsilon_{min}} \text{, } \tau=\frac{\varepsilon_{min}+\varepsilon_{max}}{2}.
\end{equation}
Now the density kernel can be calculated according to
\begin{equation}
 \mathbf{K}' \approx \mathbf{p}(\mathbf{\tilde{H}'}) = \frac{c_0}{2}\mathbf{I} + \sum_{i=1}^{n_{pl}}c_i\mathbf{T}^i(\mathbf{\tilde{H}'})
 \label{eq:chebyshev_expansion_fermi_operator}
\end{equation}
with
\begin{equation}
%  \begin{aligned}
  \tilde{\mathbf{H}}' = \mathbf{S}^{-1/2}\tilde{\mathbf{H}}\mathbf{S}^{-1/2},
%   &\tilde{\mathbf{H}}=\sigma(\mathbf{H}-\tau\mathbf{S}),\\
%   &\sigma=\frac{2}{\varepsilon_{max}-\varepsilon_{min}} \text{, }  \tau=\frac{\varepsilon_{min}+\varepsilon_{max}}{2}.
  \label{eq:chebyshev_expansion_fermi_operator2}
%  \end{aligned}
\end{equation}
where $\mathbf{I}$ is the identity matrix, $\mathbf{T}^i$ the Chebyshev polynomial of order $i$, $\mathbf{S}$ the support function overlap matrix and
$\mathbf{S}^{-1/2}$ is calculated using a first order Taylor expansion.

To determine the expansion coefficients $c_i$, one has to recall that the density matrix 
of Eq.~\eqref{eq:density_matrix} is a projection operator onto
the occupied subspace of the KS orbitals:
\begin{equation}
 \braket{\psi_i|F|\psi_j}=f(\varepsilon_j)\delta_{ij}.
 \label{eq:densitymatrix_eigenfunction_representation}
\end{equation}
Since $F$ and $\mathcal{H}$ have the same eigenfunctions, one can express the polynomial $p(\mathcal{H})$ in the same
way, leading to
\begin{equation}
 \braket{\psi_i|p(\mathcal{H})|\psi_j}=p(\varepsilon_j)\delta_{ij}
 \label{eq:chebyshevpolynomial_eigenfunction_representation}
\end{equation}
with
\begin{equation}
 p(\varepsilon)=\frac{c_0}{2}+\sum_{i=1}^{n_{pl}}c_iT^i(\varepsilon).
\end{equation}
By comparing Eqs.~(\ref{eq:densitymatrix_eigenfunction_representation}) and~(\ref{eq:chebyshevpolynomial_eigenfunction_representation}) it becomes clear that the polynomial expansion
$p(\varepsilon)$
has to approximate the Fermi function $f(\varepsilon)$ in the interval $[-1,1]$. 
Thus the coefficients $c_i$ are simply given by the expansion of the Fermi function
in terms of the Chebyshev polynomials.
The time for this step is negligible compared to
the other operations related to the FOE.
However, in practice it turns out that it is advantageous to replace the Fermi function by
\begin{equation}
 f(\varepsilon) = \frac{1}{2}\left[ 1-\text{erf} \left(\frac{\varepsilon-\mu}{\Delta\varepsilon} \right) \right],
 \label{eq:approximate_fermi_distribution_by_error_function}
\end{equation}
since it approaches the limits 0 and 1 faster as one goes away from the chemical potential. $\Delta\varepsilon$ is typically a fraction of the band gap.

The last step is to evaluate the Chebyshev polynomials and to build the density kernel.
If the $l$th column of the Chebyshev matrix $\mathbf{T}$ is denoted by $\mathbf{t}_l$, then these vectors fulfill the
recursion relation
\begin{equation}
 \begin{aligned}
  &\mathbf{t}_l^0 = \mathbf{e}_l,\\
  &\mathbf{t}_l^1 = \tilde{\mathbf{H}}'\mathbf{e}_l,\\
  &\mathbf{t}_l^{j+1} = 2\tilde{\mathbf{H}}'\mathbf{t}_l^j-\mathbf{t}_l^{j-1},
 \end{aligned}
\end{equation}
where $\mathbf{e}_l$ is the $l$th column of the identity matrix. 
The $l$th column of the density kernel, denoted by $\mathbf{k}_l$, is then given by the linear 
combination of all the columns $\mathbf{t}_l$ according to Eq.~(\ref{eq:chebyshev_expansion_fermi_operator}), i.e.\
\begin{equation}
 \mathbf{k}'_l = \frac{c_0}{2}\mathbf{t}_l^0 + \sum_{i=1}^{n_{pl}}c_i\mathbf{t}_l^i.
\end{equation}
This demonstrates that the density kernel can be constructed using only matrix vector multiplications.

Since the correct value of the Fermi energy is initially unknown, this procedure has to be repeated until the 
correct value has been found, so that $\mathrm{Tr}(\mathbf{K}')$ is equal to the number 
of electrons in the system. Finally the kernel $\mathbf{K}$ is given by $\mathbf{S}^{-1/2}\mathbf{K}'\mathbf{S}^{-1/2}$ and the
band-structure energy can then be calculated by reversing the scaling and shifting operations:
\begin{equation}
 E_{BS} = \frac{\mathrm{Tr}(\mathbf{K}\tilde{\mathbf{H}})}{\sigma} + \tau\mathrm{Tr}(\mathbf{K}\mathbf{S}).
\end{equation}

\section{Pulay forces}
\subsection{The traditional cubic approach}
\label{app:Neglect for the traditional cubic version}
Numerically, the set of $\Ket{\varPsi_i}$ is expressed in a finite basis set.
This means that the action of $\mathcal H_{KS}$ can in principle lie \textit{outside} the span of the $\Ket{\varPsi_i}$.
Let us suppose that the KS Hamiltonian and orbitals are expressed in a basis set which is complete enough to
describe them within a targeted accuracy $\Delta$. For the Daubechies basis in the traditional BigDFT approach, this happens when
the grid spacing $h$ is such as to describe the PSP and orbital oscillations, and the radii $\lambda^{c,f}$ such as to
contain the decreasing tails of the wavefunctions. This situation indeed corresponds to the traditional setup of a
BigDFT run. 
We can therefore define a residual function
\begin{equation}
 \Ket{\chi_i} = \mathcal H_{KS} \Ket{\varPsi_i} - \epsilon_i \Ket{\varPsi_i}\;,
 \label{eq:KS_residue}
\end{equation}
which is of course zero when the \textit{numerical} KS orbital is the \textit{exact} KS orbital.
By definition $\Braket{\varPsi_j | \chi_i}=0$ $\forall i,j$.
The norm of this vector, once \textit{projected} in the basis set used to express $\Ket{\varPsi_i}$, is often used as a
convergence criterion for the ground state energy.

Even though the basis set is finite, the orthogonality of the KS orbitals holds exactly,
implying $\mathrm{Re}\left(\Braket{\varPsi_i|\frac{\mathrm{d}\varPsi_i}{\mathrm{d}\mathbf{R}_a}}\right)=0$.
It is thus easy to show that the \textit{numerical} atomic forces are defined as follows:
\begin{eqnarray}
 -\frac{\mathrm{d} E_{BS}}{\mathrm{d} \mathbf R_a} = &-& \sum_i \Bra{\varPsi_i}\frac{\partial \mathcal H_{KS}}{\partial
\mathbf R_a} \Ket{\varPsi_i}\nonumber \\ &-& 2\sum_i \mathrm{Re}\left(\Braket{\chi_i| \frac{\mathrm{d} \varPsi_i}{\mathrm{d}
\mathbf R_a}} \right)\;,\label{pulaycubic}
\end{eqnarray}
where the first term of the right hand side of the above equation is the Hellmann-Feynman contribution to the forces.
The norm of $\Ket{\chi_i}$ (Eq.~\eqref{eq:KS_residue}) can be reduced within the \textit{same} basis set to meet the targeted
accuracy $\Delta$.
Therefore the projection of $\Ket{\frac{\mathrm{d} \varPsi_i}{\mathrm{d} \mathbf R_a}}$ onto the basis set used for the
calculation can be safely neglected as it is associated with the same numerical precision.
Consequently, the atomic forces can be evaluated by the Hellmann-Feynman term only as the remaining part is proportional to $\Delta$.

\subsection{The minimal basis approach}
\label{app:The case of the minimal basis setup}
As mentioned in the main text, when the KS orbitals are expressed in terms of the support functions, an additional Pulay-like term should in principle be taken into account. 
To demonstrate this, we define -- in analogy to Eq.~\eqref{eq:KS_residue} -- the support function residue $\Ket{\chi_\alpha}$, 
which becomes, using the identity $\tilde{H}_{\rho\sigma} = \sum_j c_j^\rho \varepsilon_j c_j^{\sigma *}$,
\begin{equation}
 \begin{aligned}
  \Ket{\chi_\alpha} &= \mathcal{H}_{KS}\Ket{\phi_\alpha} - \left( \sum_{\rho,\sigma}\Ket{\phi_\rho}\tilde{H}_{\rho\sigma}\Bra{\phi_\sigma} \right) \Ket{\phi_\alpha} \\
                    &= \mathcal{H}_{KS} \Ket{\phi_\alpha} - \sum_j \sum_{\rho,\sigma} c_j^\rho \varepsilon_j c_j^{\sigma *} S_{\sigma \alpha} \Ket{\phi_\rho}.
  \end{aligned}
  \label{eq:support_function_residue}
\end{equation}
Next, inserting the definition of $\chi_i$ (Eq.~\eqref{eq:KS_residue}) into the non-Hellmann-Feynman contribution of Eq.~\eqref{pulaycubic} and 
using the relation $\mathrm{Re}\left(\Braket{\varPsi_i|\frac{\mathrm{d}\varPsi_i}{\mathrm{d}\mathbf{R}_a}}\right)=0$ one obtains
\begin{equation}
\mathbf{F}_a-\mathbf{F}_a^{(HF)} = -2\sum_{i}\mathrm{Re}\left(\Braket{\varPsi_i|\mathcal{H}_{KS}|\frac{\mathrm{d}\varPsi_i}{\mathrm{d}\mathbf{R}_a}}\right).
\end{equation}
Expanding the KS orbitals in terms of the support functions, using the 
relation $H_{\alpha\beta} = \sum_j \sum_{\rho,\sigma} \varepsilon_j c_j^\rho c_j^{\sigma *} S_{\alpha\rho}S_{\sigma\beta}$ 
and the orthonormality of the KS orbitals, we can write
\begin{equation}
 \begin{aligned}
  \mathbf{F}_a-\mathbf{F}_a^{(HF)} &= -2\sum_{i}\sum_{\alpha,\beta}\mathrm{Re}\left( c_i^{\alpha *} c_i^\beta \Braket{\phi_\alpha|\mathcal{H}_{KS}|\frac{\mathrm{d}\phi_\beta}{\mathrm{d}\mathbf{R}_a}}\right) \\
                                   & \quad -2\sum_{i}\sum_{\alpha,\beta}\mathrm{Re}\left( c_i^{\alpha *} \frac{\mathrm{d}c_i^\beta}{\mathrm{d}\mathbf{R}_a} \Braket{\phi_\alpha|\mathcal{H}_{KS}|\phi_\beta}\right) \\
%                                   &= -2\sum_{i}\sum_{\alpha,\beta}\mathrm{Re}\left( c_i^\alpha c_i^\beta \Braket{\phi_\alpha|\mathcal{H}_{KS}|\frac{\mathrm{d}\phi_\beta}{\mathrm{d}\mathbf{R}_a}}\right) \\
%                                   & \quad -2\sum_{i,j}\sum_{\alpha,\beta,\rho,\sigma}\mathrm{Re}\left( c_i^\alpha \frac{\mathrm{d}c_i^\beta}{\mathrm{d}\mathbf{R}_a} \varepsilon_j c_j^\rho c_j^\sigma S_{\alpha\rho} S_{\sigma\beta} \right) \\
                                   &= -2\sum_{i}\sum_{\alpha,\beta}\mathrm{Re}\left( c_i^{\alpha *} c_i^\beta \Braket{\phi_\alpha|\mathcal{H}_{KS}|\frac{\mathrm{d}\phi_\beta}{\mathrm{d}\mathbf{R}_a}}\right) \\
                                   & \quad -2\sum_{i}\sum_{\beta,\sigma}\mathrm{Re}\left( \frac{\mathrm{d}c_i^\beta}{\mathrm{d}\mathbf{R}_a} \varepsilon_i c_i^{\sigma *} S_{\sigma\beta} \right).
 \end{aligned}
 \label{eq:force_appendix_start}
\end{equation}
From the orthonormality of the KS orbitals one can derive the relation
\begin{equation}
 2\sum_{\alpha,\beta} \mathrm{Re}\left(\frac{\mathrm{d}c_i^{\alpha}}{\mathrm{d}\mathbf{R}_a} c_i^{\beta *} S_{\alpha\beta}\right) = -\sum_{\alpha,\beta} c_i^{\alpha *} c_i^\beta \frac{\mathrm{d}S_{\alpha\beta}}{\mathrm{d}\mathbf{R}_a}.
\end{equation}
Inserting this into Eq.~\eqref{eq:force_appendix_start} yields
\begin{equation}
 \begin{aligned}
  \mathbf{F}_a-\mathbf{F}_a^{(HF)} &= -2\sum_{i}\sum_{\alpha,\beta}\mathrm{Re}\left( c_i^{\alpha *} c_i^\beta \Braket{\phi_\alpha|\mathcal{H}_{KS}|\frac{\mathrm{d}\phi_\beta}{\mathrm{d}\mathbf{R}_a}}\right) \\
                                   & \quad +\sum_{i}\sum_{\beta,\sigma}\mathrm{Re}\left( c_i^\beta c_i^{\sigma *} \frac{\mathrm{d}S_{\sigma\beta}}{\mathrm{d}\mathbf{R}_a} \varepsilon_i \right).
 \end{aligned}
\end{equation}
Again using the KS orthonormality condition, we can write
\begin{widetext}
\begin{equation}
 \begin{aligned}
  \mathbf{F}_a-\mathbf{F}_a^{(HF)} &= -2\sum_{i}\sum_{\alpha,\beta}\mathrm{Re}\left( c_i^{\alpha *} c_i^\beta \Braket{\phi_\alpha|\mathcal{H}_{KS}|\frac{\mathrm{d}\phi_\beta}{\mathrm{d}\mathbf{R}_a}}\right)
                                    +\sum_{i,j}\sum_{\alpha,\beta,\rho,\sigma}\mathrm{Re}\left( c_i^{\alpha *} c_i^\beta c_j^{\sigma *} \varepsilon_i c_j^{\rho} S_{\alpha\rho} \frac{\mathrm{d}S_{\sigma\beta}}{\mathrm{d}\mathbf{R}_a} \right) \\
                                   &= -2\sum_{\alpha,\beta}\mathrm{Re}\left( K^{\beta\alpha} \Braket{\phi_\alpha|\mathcal{H}_{KS}| \frac{\mathrm{d}\phi_\beta}{\mathrm{d}\mathbf{R}_a}}\right)
                                    +2\sum_{j}\sum_{\alpha,\beta,\rho,\sigma}\mathrm{Re}\left( K^{\beta\alpha} c_j^{\sigma *} \varepsilon_j c_j^\rho S_{\alpha\rho} \Braket{\phi_\sigma|\frac{\mathrm{d}\phi_\beta}{\mathbf{R}_a}} \right),
 \end{aligned}
\end{equation}
\end{widetext}
which becomes in terms of the support function residue of Eq.\eqref{eq:support_function_residue}
\begin{equation}
\mathbf{F}_a-\mathbf{F}_a^{(HF)} = -2\sum_{\alpha,\beta}\mathrm{Re}\left( K^{\beta\alpha} \Braket{\chi_\alpha|\frac{\mathrm{d}\phi_\beta}{\mathrm{d}\mathbf{R}_a}} \right).
\end{equation}
%and the non Hellmann-Feynman term can -- as shwon in more detail in Appendix~\ref{app:derivation_forceformula} -- be written as follows:
% \begin{equation}
%  \mathbf F_a - \mathbf F^{(HF)}_a = -2 \sum _{\alpha, \beta}\mathrm{Re}\left( K^{\alpha \beta}\right)
% \Braket{\chi_\beta| \frac{\mathrm{d} \phi_\alpha}{\mathrm{d} \mathbf R_a}} \;.
% \end{equation}
This result contains Eq.~\eqref{pulaycubic} when no localization projectors are
applied to the support function. Therefore the only term of the forces which cannot be captured within the localization
regions is the part which is projected outside. The extra Pulay term due to the localization
constraint is therefore 
\begin{equation}
 \mathbf F^{(P)}_a=-2 \sum _{\alpha, \beta}\mathrm{Re}\left( K^{\beta\alpha } \Braket{ \chi_\alpha|(1 - \mathcal
L^{(\beta)}) |\frac{\mathrm{d}
\phi_\beta}{\mathrm{d} \mathbf R_a}} \right)\;.
\end{equation}
Using Eq.~\eqref{externalderivative}, we can show
\begin{equation}
\mathbf F^{(P)}_a=
-2 \sum _{\alpha, \beta}\mathrm{Re}\left( K^{\beta\alpha } \Braket{\chi_\alpha |\frac{\partial \mathcal
L^{(\beta)}}{\partial \mathbf R_a} |\phi_\beta}\right)\;.
\end{equation}
When the localization regions are atom-centered, the derivative of the projector
$\mathcal L^{(a)}$ (as defined in Eq.~\eqref{eq:loccon}) can be evaluated analytically in the underlying basis set and is given by
\begin{eqnarray}\label{eq:locprojderiv}
 \frac{\partial \mathcal L^{(\alpha)}}{\partial \mathbf R_\beta}_{i_1,i_2,i_3;j_1,j_2,j_3} = &&\delta_{\alpha\beta}
\delta_{i_1j_1} \delta_{i_2
j_2} \delta_{i_3j_3}\frac{\mathbf{R}_{(i_1,i_2,i_3)}-\mathbf{R}_\alpha}{R_{cut}} \nonumber \\
&\times&
\delta(R_{cut} -|\mathbf{R}_{(i_1,i_2,i_3)}-\mathbf{R}_\alpha|)
\;.
\end{eqnarray}
This demonstrates that the Pulay term is only associated with the value of the
support functions at the border of the localization regions.

\section{Parallelization}
\label{app:parallelization}

It is a natural choice to divide the support functions between MPI tasks so that each one
handles only a subset of support functions. For some operations these can be treated independently but for 
others, such as the calculation of scalar products between overlapping support functions needed to build 
the overlap and Hamiltonian matrices, communication of support functions between MPI tasks is required.  
One could directly exchange in a point-to-point 
fashion the parts of the support functions which overlap with each other, so that the scalar products can be calculated
locally on each task. Although conceptually straightforward, this has severe drawbacks.  
Firstly the amount of data being communicated is tremendous since the support functions generally have quite a notable
overlap.  This also results in a very poor ratio between computation and communication -- in the extreme case 
 where each task handles only one support function, each communicated element is only used
for one operation. 
Secondly there can be enormous load imbalancing for free boundary conditions as support functions in the center of the 
system usally have more neighboring support functions than those near the edges. 
Finally, the data is split into a large number of small messages, which could result in a large overhead due to 
the latency of the network.

\begin{figure}
 \centering
 \includegraphics[width=0.48\textwidth]{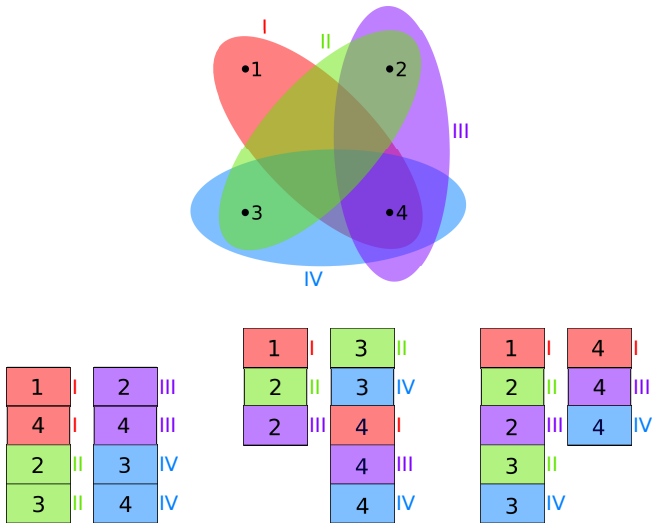}
 \caption{An example depicting four support functions (Roman numerals) in a system 
	 consisting of four grid points (Arabic numerals) [above]. 
 The support functions are constructed such that each extends over two grid points.
 The various data layouts are also illustrated [below]: 
 the direct layout where each MPI task has all the data for certain support functions [left]; 
 the naive transposed layout where each MPI task has all the data for some grid points [centre]; 
 and the optimized transposed layout which is similar to the previous case, but with an optimal load balancing [right].}
 \label{fig:example_orbitals_together}
\end{figure}

We therefore use a different approach, which requires a so-called ``transposed'' rather than ``direct'' 
arrangement of data. In this layout the simulation cell is partitioned among MPI tasks and
the support functions are distributed to the various tasks such that 
each one can calculate a partial overlap matrix for a given region of the cell. Each task therefore has to receive those parts of all
support functions which extend into its region. 
The partial matrices are then summed to build the full overlap matrix using MPI\_Allreduce. This partitioning of the 
cell is done such that the load balancing among the MPI tasks is optimal, which in general does not correspond to a
naive uniform distribution of the simulation cell. 
To determine the optimal layout a weight is assigned to each grid point, given by $m^2$, where $m$ is the number of 
support functions touching it (if symmetry can be exploited the weight should rather be $\frac{m(m+1)}{2}$); 
the total weight (i.e. the sum of all partial weights) is then divided among all MPI tasks as evenly as possible.
In Fig.~\ref{fig:example_orbitals_together} this procedure is illustrated with a toy example, 
where in the upper part the support functions and their overlaps are shown and in the lower part 
the resulting direct and transposed (both naive and optimal) data layouts are given.

\begin{figure}
 \centering
 \includegraphics[width=.48\textwidth]{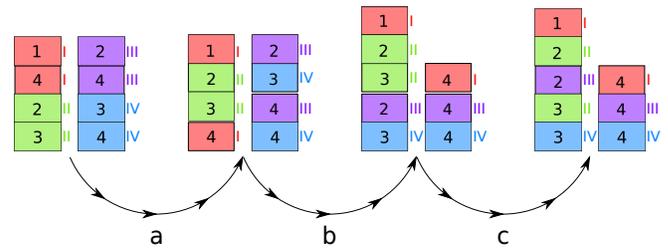}
 \caption{Illustration of the transposition process for the system shown in Fig.~\ref{fig:example_orbitals_together}. 
In step a, the data is rearranged locally on each MPI task, after which it is communicated using a single collective call
(MPI\_Alltoallv), as shown in step b. Finally in step c it is again rearranged locally to reach the final layout.}
 \label{fig:transposition_process}
\end{figure}

In addition to the better load balancing this approach has the advantage that considerably less data has to 
be communicated -- since the transposed layout is just a redistribution 
of the standard layout, the total amount of data that is communicated is equal to the total size of all support
functions, whereas in the point-to-point approach, the same data is often sent to multiple processes.
Furthermore, the communication can be done more efficiently: after some local rearrangement of 
the data for each MPI task, it can be communicated with a single MPI call (MPI\_Alltoallv) -- 
in practice there are two calls since the coarse and fine parts are handled separately. After the data has 
been received some local rearrangement is again required to reach the correct layout. 
These three steps -- local rearrangement, communication and further local 
rearrangement -- are illustrated in Fig.~\ref{fig:transposition_process}. Due to the latency of the network, 
two MPI calls will likely be more efficient than 
the very large number of small messages that have to be sent for the point-to-point approach.

For the calculation of the charge density, which is formally identical to the calculation of scalar products, 
a similar approach is used. Since these two operations are the most important ones from the viewpoint of
 communication and parallelization, this results in an excellent scaling with respect to the
number of cores.

\bibliography{base}

\end{document}